\documentclass[pre,superscriptaddress,twocolumn,letterpaper]{revtex4-1}

\usepackage{graphicx}
\usepackage{amssymb}
\usepackage{chemarr}
\usepackage{appendix}
\usepackage{setspace}
\usepackage{xcolor}
\usepackage{soul}
\DeclareMathOperator{\sign}{sgn}

\begin{document}

\title{Mechanisms of Cooperation and Competition of two Species Transport in Narrow Nanochannels }

\author{Wolfgang Rudolf Bauer\thanks{corresponding author\\
Tel. +49-931-201-39011\\}}
\email{bauer\_w@ukw.de}
\affiliation{
Comprehensive Heart Failure Center, Am Schwarzenberg 15, A15 \\ \& \\ Dept. of Internal Medicine I, University Hospital of W\"urzburg,
Oberd\"urrbacher Stra{\ss}e 6,
D-97080 W\"urzburg, Germany
}


\date{\today}


\begin{abstract}
Flow of particles of two different species through a narrow channel with solely two discrete spatial positions is analyzed with respect to the species' capability to cooperate or compete for transport. Besides blocking its own position within the channel also interparticle interactions between neighboring particles in the channel are considered. The variety of occupation options within the channel defines the state space. The transition dynamics within is considered as a continuous Markov process. So, in contrast to mean field approaches, spatial correlations are explicitly conserved. A strong repulsive interaction between particles of the same kind and a very attractive channel imply a strong entanglement of transport of both species. This is reflected by the magnitude of transition flows in state space which in the extreme case of perfect coupling are restricted to a cyclic sub space. Entanglement of transport implies that the species mutually exert entropic forces on each other. For parallel directed concentration gradients this implies that the species' ability to cooperate increases with the degree of entanglement. Thus, the gradient of one species reciprocally induce a higher flow of the other species when compared to that in its absence. The opposite holds for antiparallel gradients where species mutually hamper their transport. For a sufficient strong coupling, the species under the influence of the stronger concentration gradient drives the other against its gradient, i.e. flow and gradient of the driven species become antiparallel. Hence, besides the positive entropy production generated by the driving species a negative component of entropy production of the driven species emerges. The sources of both, positive and negative entropy production, can be localized in state space. The stronger the coupling of transport the higher is the degree of efficiency, i.e. the amount of negative entropy production on cost of the positive one.             
\end{abstract}
\maketitle

\section{Introduction}
Particle channel transport and its regulation is of paramount importance in biological systems as well as in applications in nanotechnology \cite{Iqbal,Talisman2009nat}. Besides this the conceptual framework describing channel transport may be easily extended also to describe non-spatial \"transport\", e.g. along the reaction coordinate of enzymatic reactions.  

The transport itself depends on thermodynamic forces, e.g. concentration gradients between the domains connected by the channel, electrical drift forces as well as on particle-channel -, and interparticle interactions. When more than one species is involved  inter- and intraspecies interactions must be differentiated.   

An intriguing question of mixed species transport is, under which circumstances particles of different species may cooperate, or  mutually on cost of the own species promote the other one, or solely compete for transport. This sophisticated behaviour could be recently shown by us in a simple Markovian model of channel transport \cite{Bauer2013}, which, in contrast to mean field models, explicitly conserved spatial correlations of interparticle interactions, and was numerically exactly solvable. 
We could demonstrate that all capabilities of cooperative and competitive interactions between the two species increase with the length of the channel. This lead us to the hypothesis that the amount of interspecies interactions within the channel is the crucial factor, as longer channels offer more options for particles of different species to interact. 

In this paper we will focus in detail on the impact of inter- and intra-species interaction on cooperation and competition. We will do this in the most minimalist model of a channel, which maintains intra- and interspecies interaction inside, i.e. a spatial discrete model with solely two positions inside the channel. 

The manuscript is structured as follows: first we give an outline of the model, in which we introduce its state space, its Markovian transition dynamics, and the resulting implications for probability and particle flow in the steady state. In the second section we consider stochastic paths in state space the system has to undergo to realize particle transport between the baths. We identify which states the paths must contain and which not to achieve an optimum coupling of transport of the two species, which defines the relevant interactions. The implications of an increasing entanglement of the species' transport on flow in state space, and particle flow between the baths is shown and we analyze the impact on the species' capability to cooperate, promote or compete. In the 3rd section entropy production of the species and its sources in state space are analyzed as a function of the coupling strength, i.e. degree of entanglement of both transport processes.    

\section{The model}
\begin{figure*}
\includegraphics[trim = 1mm 1mm 1mm 1mm, clip,width=16cm]{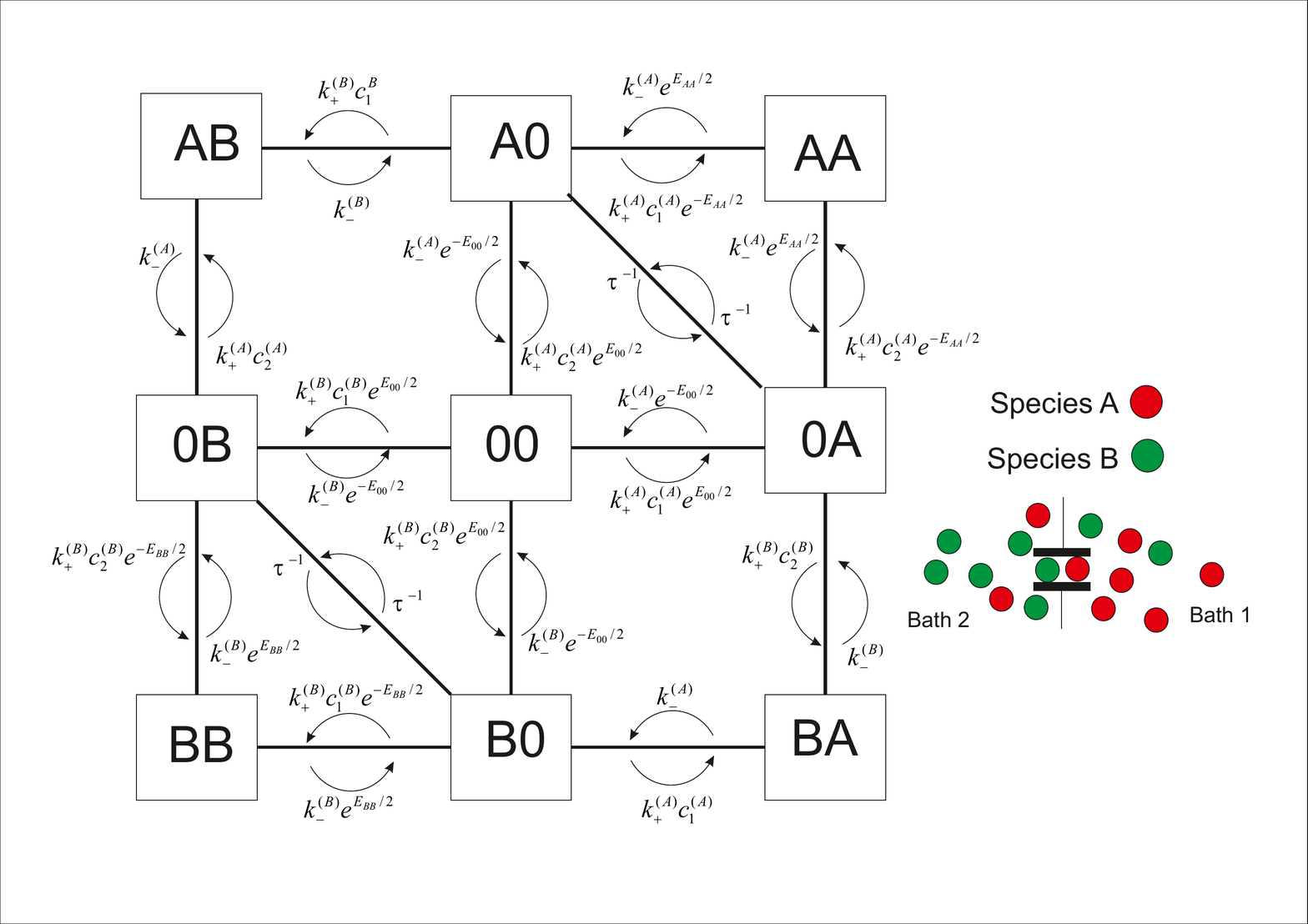}
\caption{The 9-dim state space $\boldsymbol{\Sigma}=\lbrace\boldsymbol{\sigma}| \boldsymbol{\sigma}=(\sigma_2,\sigma_1),\;\sigma_i=0,A,B\rbrace$, with transition rates between states. Right: a sketch of the channel connecting the two baths.} 
\label{statespace}
\end{figure*}

Our channel model was recently described in detail \cite{Bauer2013}. Briefly:  
The channel connects two baths  $1$ and $2$, with  particles of two species $X=A,\;B$ with respective concentrations $(c_1^{(X)},c_2^{(X)})$ inside. 
Our minimalist model shall only allow two spatial positions for particles inside the channel, each of which may at most be occupied  by only one particle. This implies that any particle transition to a position in the channel demands its vacancy.  
A channel state is completely described by the state variable $\boldsymbol{\sigma}=(\sigma_2,\;\sigma_1)$, where the values of $\sigma_i$ indicate whether the position $i$ is empty ($\sigma_i=0$) or occupied by a particle of species $A$ ($\sigma_i=A$) or $B$ ($\sigma_i=B$). These states form a $3^2=9$-dimensional state space $\boldsymbol{\Sigma}$ (see Fig.~\ref{statespace}).

For simplicity the particle-channel interactions is considered to be homogeneous, i.e. its profile inside  the channel is flat. This implies that the transition rates between the two positions are equal, i.e. $r_{(X,0)\to(0,X)}=r_{(X,0)\leftarrow(0,X)}$. In addition we assume the rates to be same for the two species. This rate $r$, which is a measure of the particles mobility in the channel, defines the time constant $\tau=r^{-1}$, to which we normalized all temporal parameters, i.e. 
\begin{equation}
\tau=r^{-1}=1\;.\label{transition1}
\end{equation}
Further we assume symmetric exchange dynamics of particles at the channel ends with the respective bahts. This, and the flat particle-channel interaction profile, imply that particle transport is merely driven by concentration gradients, and not by any energetic potential differences between the baths. Hence the free energy gain of particles, when passing from one bath to the other is determined from the difference of the chemical potentials as \footnote{we consider the gain of free energy as the negative of the corresponding free energy difference $\Delta\epsilon$, i.e. in this case $\Delta\epsilon=-\Delta\mu$}.   
\begin{equation}
\Delta\mu^{(X)}=\ln\big(c_1^{(X)}/c_2^{(X)}\big)\;.
\end{equation}

Transitions from and to a bath are restricted to respective adjacent channel positions. If we had solely the blocking interparticle interaction, the transition rate describing the access dynamics from the bath to a vacant channel position would proportional to some rate constant $k_+$ times the particle concentration in the respective bath. Vice versa particles would leave such a position toward the adjacent bath with a rate $k_-$, where the potential difference $\Delta\Phi=-\ln(k_+/k_-)$ describes the binding strength the channel exerts on the particle \footnote{Note that we normalize all energetic quantities to $kT$, i.e. $\Phi\to \Phi/kT$, with $k$ as the Boltzmann constant and $T$ as the temperature}. The access of particles from the bath solely requires an empty adjancent channel position and would be independent from the occupation state of the non-adjacent site. However, we now want to consider a more sophisticated interparticle interaction than simple blocking. This can be realized by energetic differences, influencing the bath-channel exchange dynamics, which depend on the occupation state of non-adjacent channel site. So we introduce an energetic difference $E_{X,X}$ a particle has to pass from the bath to a vacant spatial position if the channel is already occupied by a particle of the same species $X$, which would e.g. hold for a transition $(0A)\to(AA)$. If $E_{AA}>0$ this acts as a repulsive interaction between particles of the same species. The corresponding rates for transitions e.g. at the left side of the channel then become \cite{Nadler86}
\begin{eqnarray}
r_{(0X)\to (XX)}&=& k_+ e^{-E_{XX}/2}\; c_2^{(X)}\;,\cr
r_{(XX)\to (0X)}&=& k_- e^{E_{XX}/2} \;.\label{transition2}
\end{eqnarray}
The same holds symmetrically for right side of the channel. Bath-channel transitions of particles which enter a channel occupied by a particle of a different species shall solely be described by the rate constants $k_+,\;k_-$.     
We also introduce an energy difference $E_{00}$ describing the affinity of the empty channel ($\boldsymbol\sigma=(00)$) to absorb a particle, i.e. this energy is gained when a particle enters an empty channel. These rates are \cite{Nadler86}
\begin{eqnarray}
r_{(0,0)\to \hbox{one particle in channel}}&=& k_+ e^{E_{00}/2}\; c_{X}\;,\cr
r_{\hbox{one particle in channel}\to (0,0)}&=& k_- e^{-E_{00}/2} \;. 
\label{transition3} 
\end{eqnarray}
This affinity of the empty channel is assumed to be identical for the two species. 

Transition dynamics on the state space is that of a continuous stationary Markov process. The evolution of the probabilities 
$\boldsymbol{P}=(P_{\boldsymbol{\sigma}}(t))_{\boldsymbol{\sigma}\in \boldsymbol{\Sigma}}$ to find the channel in the respective states is then determined by a Master Equation
\begin{equation}
\frac{d}{dt}\;\boldsymbol{P}(t)=\boldsymbol\Lambda\;\boldsymbol{P}(t)\;, \label{MasterEq}
\end{equation}
with the $3\times 3$ matrix $\boldsymbol\Lambda=(\lambda_{\boldsymbol{\sigma},\boldsymbol{\varsigma}})$ containing  the transition rates $\lambda_{\boldsymbol{\sigma},\boldsymbol{\varsigma}}=\lambda_{\boldsymbol{\sigma}\leftarrow \boldsymbol{\varsigma}}$ from channel states $\boldsymbol{\varsigma}$ to  $\boldsymbol{\sigma}$ \footnote{we notate the sequence  of state variables of a transition $\boldsymbol{\sigma}\leftarrow \boldsymbol{\varsigma}$ as $\boldsymbol{\sigma},\boldsymbol{\varsigma}$ in the index of the transition rate $\lambda$ to be in line with the usual index notation in matrix algebra}. They are given by Eqs.~(\ref{transition1}-\ref{transition3}), and can be depicted from Fig.~(\ref{statespace}). 

As the system must be in some state, conservation of probability holds, i.e. $d/dt\sum_{\boldsymbol{\sigma}} P_{\boldsymbol{\sigma}}=0$. This determines the diagonal matrix elements of $\boldsymbol\Lambda$ as  
\begin{equation}
\lambda_{\boldsymbol{\varsigma},\boldsymbol{\varsigma}}=-\sum_{\substack{\boldsymbol{\sigma}\in\boldsymbol{\Sigma}\\ \boldsymbol{\sigma}\neq\boldsymbol{\varsigma}}}\lambda_{\boldsymbol{\sigma},\boldsymbol{\varsigma}}\;. \label{DiagonalE}
\end{equation}   

The transition rates between the states $\boldsymbol{\sigma}\leftrightarrows \boldsymbol{\varsigma}$ define a free energy difference 
\begin{equation}
\Delta\epsilon_{\boldsymbol{\sigma},\boldsymbol{\varsigma}}=-\ln\left(\frac{\lambda_{\boldsymbol{\sigma},\boldsymbol{\varsigma}}}{\lambda_{\boldsymbol{\varsigma},\boldsymbol{\sigma}}}\right)\;,\label{freenergydiff}
\end{equation}
which results either from energetic differences and/or that of entropic forces related to particle exchange. This free energy differences acts as the driving force for the net flow of probability, which is given by 
\begin{equation}
J_{\boldsymbol{\sigma} \leftarrow \boldsymbol{\varsigma}}=J_{\boldsymbol{\sigma}, \boldsymbol{\varsigma}}=\lambda_{\boldsymbol{\sigma},\boldsymbol{\varsigma}} P_{\boldsymbol{\varsigma}}-\lambda_{\boldsymbol{\varsigma},\boldsymbol{\sigma}}P_{\boldsymbol{\sigma}}\;.\label{flows}
\end{equation}
With Eq.~(\ref{DiagonalE}) one can rewrite the Master Equation (\ref{MasterEq}) in the form of continuity equation of probability 
\begin{equation}
\frac{d}{dt}\;P_{\boldsymbol{\sigma}}=\sum_{\boldsymbol{\varsigma}\in \boldsymbol{\Sigma}} J_{\boldsymbol{\sigma}, \boldsymbol{\varsigma}}\;. \label{conservationp}
\end{equation}
The latter describes that the change of probability to find the system in state $\boldsymbol{\sigma}$ results from the probability flows directed to it from all other states $\boldsymbol{\varsigma}$.
  
In the absence of particle concentration gradients between the baths the system is in thermodynamic equilibrium and detailed balance holds, i.e. all flows between states in Eq.~(\ref{flows}) vanish. Thus, the equilibrium probabilities $P^{(e)}$ fulfill 
\begin{equation}
P_{\boldsymbol{\sigma}}^{(e)}/P_{\boldsymbol{\varsigma}}^{(e)}=\lambda_{\boldsymbol{\sigma},\boldsymbol{\varsigma}}/\lambda_{\boldsymbol{\varsigma},\boldsymbol{\sigma}}=e^{-\Delta\epsilon_{\boldsymbol{\sigma},\boldsymbol{\varsigma}}^{(e)}}\;.
\end{equation}
In this case we can assign the states a potential 
\begin{equation}
\phi_{\boldsymbol{\sigma}}=-\ln(P_{\boldsymbol{\sigma}}^{(e)})\;\label{equilibrium}
\end{equation}
i.e. the free energy differences between states are that between the corresponding potentials 
\begin{equation}
\Delta\epsilon_{\boldsymbol{\sigma},\boldsymbol{\varsigma}}^{(e)}=\phi_{\boldsymbol{\sigma}}-\phi_{\boldsymbol{\varsigma}}\label{potential}
\end{equation}
This implies that we have a conservative field of driving forces in the state space, i.e. the free energy difference along a path $[\hbox{start}=\boldsymbol{\sigma}_1\cdots\boldsymbol{\sigma}_N=\hbox{end}] $ in state space, 
\begin{equation}
\sum_{i=1}^{N}\Delta\epsilon_{\boldsymbol{\sigma}_{i+1},\boldsymbol{\sigma}_i}^{(e)}=\phi_{\hbox{\tiny{end}}}-\phi_{\hbox{\tiny{start}}}\;,
\end{equation}
solely depends on the start and end state of the path. In particular it vanishes for closed paths. 

Non-vanishing concentration gradients of particles between the baths induce a non-conservative field of driving forces on state space, i.e. there exist closed paths in state space in which free energy is gained related to particle transport between the baths. For example the closed path $[0A - A0 - 00 - 0A]$ describes particle transport of species $A$ from bath~1 to bath~2 for which we gain the free energy $\ln(c_1^{(A)}/c_2^{(A)})$. In the following we want to restrict to stationary non-equilibrium conditions, i.e. the system is in a the steady state, implying that its probability distribution $\boldsymbol{P}_s$ remains constant in time. Hence, 
\begin{equation}
\frac{d}{dt}\;\boldsymbol{P}_s(t)=\boldsymbol\Lambda\;\boldsymbol{P}_s \equiv  0\;, \label{steadystate}
\end{equation}
holds, which, with Eq.~(\ref{conservationp}), implies the conservation of flow around any state $\boldsymbol{\sigma}$ (Kirchhoff's cicuit laws)
\begin{equation}
\sum_{\boldsymbol{\varsigma}\in \boldsymbol{\Sigma}} J_{\boldsymbol{\sigma}, \boldsymbol{\varsigma}} =0\;\label{Kirchoff}
\end{equation}  

Particle flow between the baths depends on that at the channel ends. Flow, e.g. of species $A$, at the left channel (see Fig.~{\ref{statespace})) results from  transitions between the states $(0,\sigma)\leftrightarrows (A,\sigma) $. Hence,   
\begin{equation}
J^{(A)}=J_{(0A),(AA)}+J_{(00),(A0)}+J_{(0,B),(AB)}\;.
\end{equation}
The conservation of flow  in the steady state (see Eq.~\eqref{Kirchoff}) then implies that this flow must equal to that within the channel, $J_{(A0),(0A)}$, and that at the right channel end $J_{(AA),(A0)}+J_{(0A),(00)}+J_{(BA),(B0)}$ . The same holds for species $B$. Hence, we obtain for particle flow between the channels  
\begin{equation}
J^{(A)}=J_{(A0),(0A)}\;\hbox{and}\;J^{(B)}=J_{(B0),(0B)}\;.\label{Pflow}
\end{equation}

In the following we determine the steady state probabilities from Eq.~(\ref{steadystate}) numerically and by this flows between the states (Eq.~(\ref{flows})) and of particles through the channel (Eq.~(\ref{Pflow}).

\section{Cooperation and Competition}

\begin{figure}
\includegraphics[width=8cm]{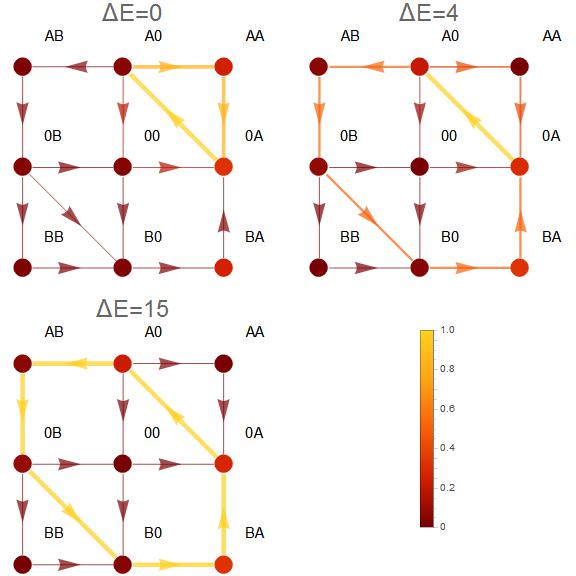}
\caption{Flows and occupation probabilities in state space. Different energetic levels of the empty channel state and channel states occupied by two particles of the same species are considered. For simplicity all are set equal to $\Delta E=E_{00}=E_{AA}=E_{BB}$.
 Values for probabilities are color coded (see bar). So are flow values, which were normalized to that with the maximum magnitude. In addition flow magnitude is coded by the thickness of the arrows, which indicate the flow direction. Particle concentrations of species $A$ in the right (1)/left (2) bath  are $c_1^{(A)}=10$ and $c_2^{(A)}=0.1$ respectively. $k_+^{(A)}$ and $k_-^{(A)}$ were set equal to 1. Concentrations of $B$ were set equal in both baths, and jump in rates were chosen to be $k_+^{(B)} c_i^{(B)}=e^{1}\times 0.1$, $i=1,\;2$ and jump out $k_-^{(B)}=e^{-1}$. This choice of rates implied a moderate attractive particle-channel interaction $\Delta \Phi^{(B)}=-\ln(k_+^{(B)}/k_-^{(B)})=-2$ for species $B$ compared with that of $A$, $\Delta \Phi^{(A)}=-\ln(k_+^{(A)}/k_-^{(A)})=0$. Note, that with increasing $\Delta E$, flow is mainly present (yellow) on the cyclic state space $CS$ (\ref{limitcycle}). } 
\label{Flowstsp}
\end{figure}

\begin{figure}
\includegraphics[width=8cm]{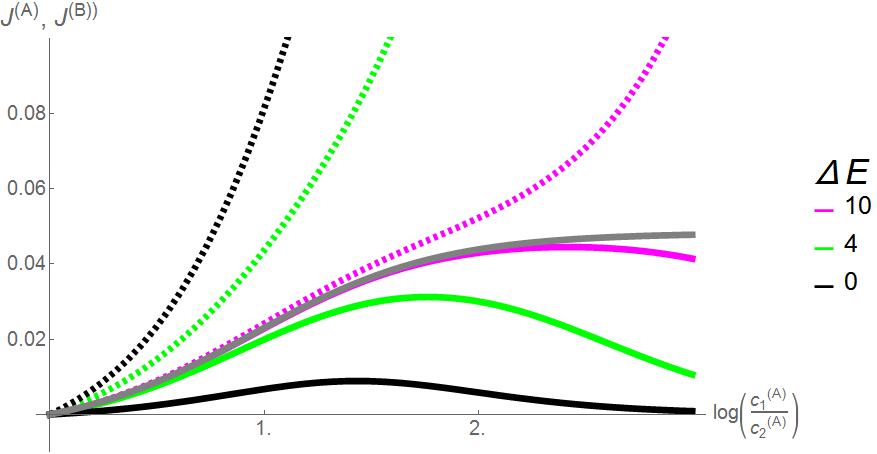}
\caption{Particle flow of species $A$ (dashed lines) and $B$ (solid lines) through the channel as a function of the concentration gradient of $A$. Particle concentration in bath~1 is elevated, that in bath~2 is held constant $k_+^{(A)} c_2^{(A)}=0.1$). Concentrations of $B$ are equal in both baths. These and the constants $k_+,\;k_-$ for both species are identical with that in Fig.~(\ref{Flowstsp}). The interaction between the respective species is varied by the energetic levels of the empty channel and of double occupied channel states of the same species. The gray line gives the flow in the restricted cyclic state space CS (\ref{limitcycle}), and is obtained from Eq.~(\ref{flowcs}). Note the flows of $A$ and $B$ converge towards this flow with increasing $\Delta E$.} 
\label{Couplingflow}
\end{figure}

Within the state-space our system undergoes stochastic transitions according to the Master equation (\ref{MasterEq}). The net thermodynamic driving forces for particle transport across the channel are the concentration gradients of particles between the respective baths. A non-vanishing net particle transport of species $X$ ($X=A,\;B$) from e.g. bath~1 to bath~2  requires the repetitive visit of the states $(0X)$ and $(X0)$ (s.~Eq.~(\ref{Pflow})). So the stochastic path of successive states may be build up from closed paths which contain the segment $(0X)-(X0)$. The entanglement of the species' transport should mainly depend on the options particles have to interact within the channel. In order to realize this interaction, our minimalist channel model with only two spatial positions inside offers solely two states, $(AB)$ and $(BA)$. Hence, interparticle interactions and particle channel interactions which favor visits to these states, or, vice versa, which hamper access to states that are not involved in paths leading to these two states, should favor cooperation or competition. So, entanglement of different species' transport is realized on closed paths which contain the segments $(0A)-(A0)$ and $(0B)-(B0)$ as well as the states $(AB)$ and $(BA)$. However, cooperation and competition do not merely depend on the presence of states in which different species coexist in the channel. Instead, being in these states, the species must mutually exert some force on each other. Here, it is an entropic force which results from the left/right bias of occupation inside the channel, which itself results from the concentration gradients. 

 The 2nd law of thermodynamics favors  
paths in which free energy is gained. In the steady state this translates into the direction and magnitude of stationary flows between states (see Eq.~\ref{flows}). From this flow pattern one can infer which paths are favored. In Fig.~(\ref{Flowstsp}) a concentration gradient drives particles of species $A$ from bath~1 to bath~2. Concentration of species $B$   was chosen to be equal in both baths, i.e. transport of the latter solely depends on its interaction with species $A$. For the native set up, i.e. when there is indifferent affinity of the empty channel to attract particles $E_{00}=0$, nor a repulsive interaction impeding occupation by particles of the same species, $E_{AA}=E_{BB}=0$, the most favored cyclic path, on which the free energy $\ln(c_1^{(A)}/c_2^{(A)})$ is gained, is $(0A)-(A0)-(AA)-(0A){\color{gray}{-(0A)\cdots}}$. Note that 
this free energy is the sum over those gained at each transition, i.e.  $\ln(c_1^{(A)}/c_2^{(A)})=-(\Delta\epsilon_{(A0),(0A)}+\Delta\epsilon_{(AA),(A0)}+\Delta\epsilon_{(0A),(AA)})$.
Only a negligible fraction of flow passes through the states $(AB)$, $(BA)$, and, hence, particle flow of $B$, which according to  Eq.~\eqref{Pflow} is identical with that from $(0B)$ to $(B0)$, is only moderate (see Fig.~(\ref{Flowstsp}) for $\Delta E=0$ and black solid line in Fig.~(\ref{Couplingflow})). Figure~(\ref{Flowstsp}) shows that with increasing affinity of the empty channel   and repulsive interactions between particles of the same species, closed paths including the states $(AB)$ and $(BA)$ become favorable. As a result the flow of the driven species $B$ increases whereas that of the driving species $A$ decreases (Fig.~\ref{Couplingflow}). In the limiting case ($E_{00}=E_{AA}=E_{BB}=\Delta E \to \infty$) visitations of states, and, hence flow in between them, is reduced to the cyclic state space $CS$ 
\begin{equation}
(0A)-(A0)-(AB)-(0B)-(B0)-(BA)-(0A){\color{gray}{-(0A)\cdots}}\;.\label{limitcycle}
\end{equation} 
As this cyclic sub state-space has no branching, all flows between connected states are equal in the steady state. In particular particle flows of species $A$ and $B$ are equal in this limiting case, i.e. 
\begin{equation}
\lim_{\Delta E\to\infty}J^{(A)}(\Delta E)=\lim_{\Delta E\to\infty}J^{(B)}(\Delta E)=J^{\hbox{cs}}\;.   \label{toflowcs}
\end{equation}             
This is shown in Fig.~(\ref{Couplingflow}). Here, with increasing $\Delta E$, the flows of both species converge towards  $J^{\hbox{cs}}$. In general this steady state flow of a circular Markov process is obtained as (see Appendix A)    
\begin{equation}
J^{\hbox{cs}}=\frac{1-e^{\Delta U}}{\tau_{+}+e^{\Delta U}\;\tau_{-} +R}\;\label{flowcs}
\end{equation}
with $\Delta U$ as the free energy difference the system experiences after one turn in the $CS$. It is obtained from the single free energy differences (Eq.~\eqref{freenergydiff}) between successive states $\boldsymbol{\sigma}_{i+1},\;\boldsymbol{\sigma}_{i}$ in CS (see scheme (\ref{limitcycle})) as 
\begin{eqnarray}
\Delta U&=&\sum_{i=1}^{5}\epsilon_{\boldsymbol{\sigma}_{i+1},\boldsymbol{\sigma}_{i}}=-\ln \left(\frac{c_1^{(A)}}{c_2^{(A)}}\right)-\ln\left(\frac{c_1^{(B)}}{c_2^{(B)}}\right)\cr\cr
&=&-\Delta\mu^{(A)}-\Delta\mu^{(B)}\;,\label{freeenergycs}
\end{eqnarray}
i.e. $|\Delta U|$ is the sum of the chemical potential differences of both species. 
$\tau_{+/-}$ are the mean first passage times the system needs to pass in (counter)clockwise(+)/- direction one turn in the CS to from arbitrary state $\boldsymbol{\sigma}_i\in CS$ \footnote{For  counterclockwise/clockwise direction select the state $\boldsymbol{\sigma}_{i+1}$ / $\boldsymbol{\sigma}_{i-1}$ as a start and impede transitions from this state in clockwise/counterclockwise direction (reflective boundary conditions). $\tau_+/-$ is then the mean time to reach $\boldsymbol{\sigma}_i$ for the first time (absorptive boundary conditions)}. $R$ is the conductivity of probability flow on CS. For explicit determination of the steady state flow in CS, first passage times and conductivity  see Appendix A.  

\begin{figure}
\includegraphics[width=8cm]{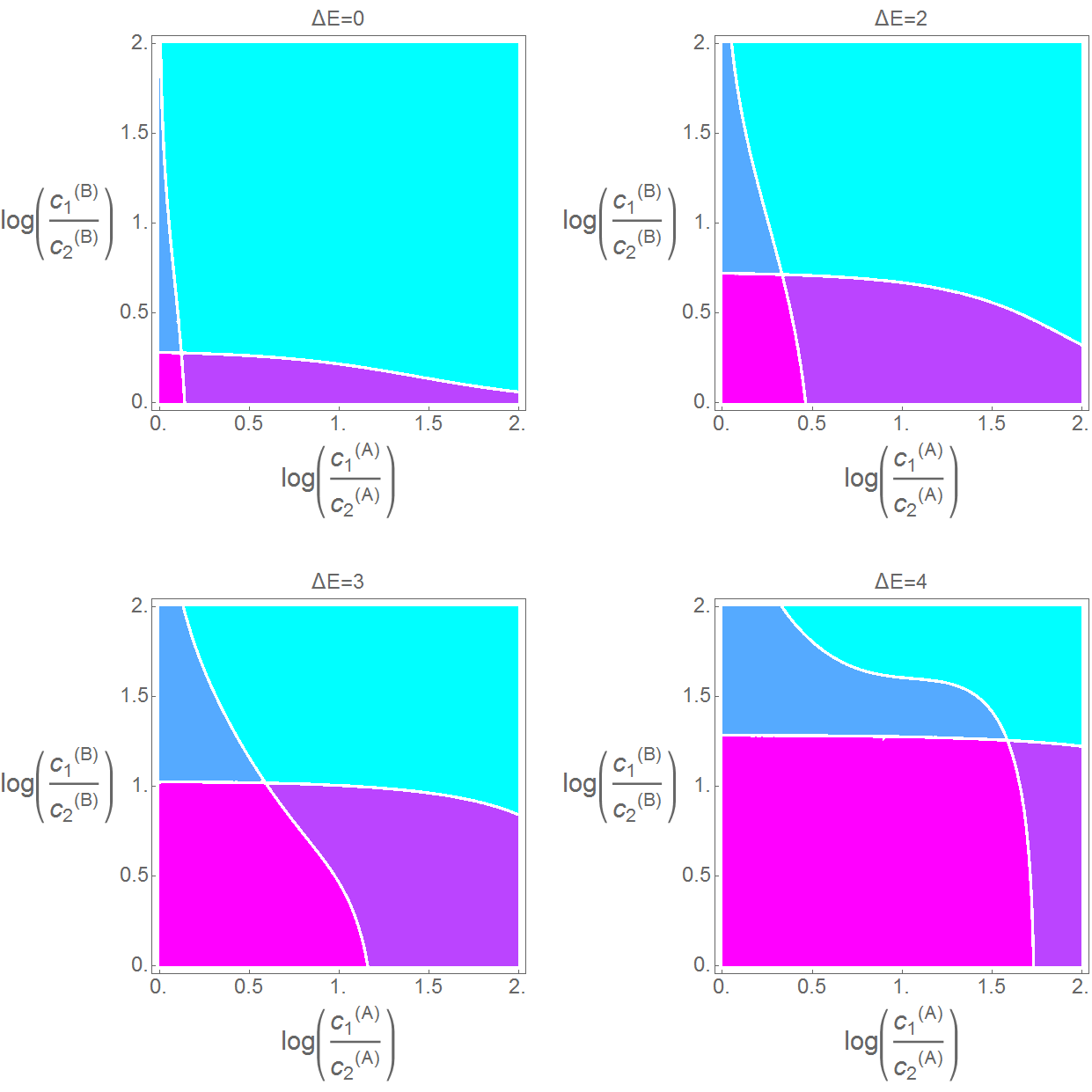}
\caption{Cooperation, promotion and competition of two species particle transport as function of their coupling, quantified by $\Delta E$. This energy difference was chosen for the empty channel affinity and the repulsive interaction of particles of the same species. Pink denotes cooperation with profit for both, turquoise competition on cost of both, violet species $B$ is promoted by $A$ on cost of the latter, and vice versa blue (see text).} 
\label{Phasediagram}
\end{figure}

\begin{figure}
\includegraphics[width=8cm]{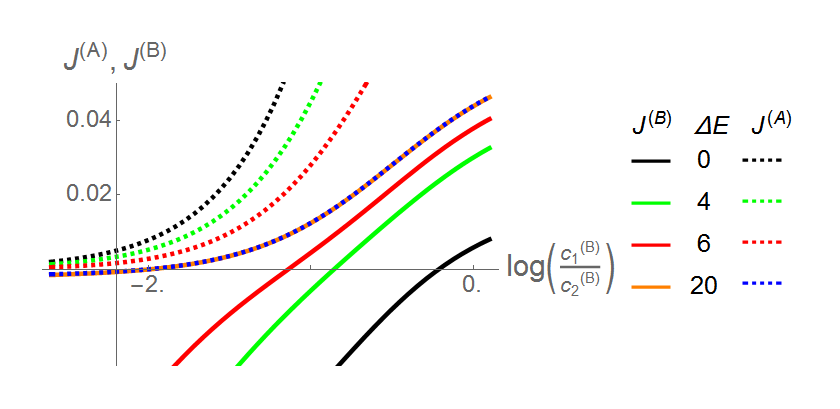}
\caption{Effect of opposing concentration gradient of species $A$ and $B$ on respective flows as a function of the coupling strength $\Delta E$. Concentrations and jump in and out rates for species A are that of Fig.(\ref{Flowstsp}), i.e. the concentration gradient is directed from bath~1 to bath~2, with $c_1^{(A)}/c_2^{(A)}=100$. Concentration of $B$ in bath~1 is held constant at $c_1^{(B)}=0.1$, and $c_2^{(B)}$ is increased. The stronger the coupling the higher must be the gradient of $B$ to make its flow cease.} 
\label{Opposingrad}
\end{figure}

In CS there is maximum entanglement of transport of both species. Equation (\ref{toflowcs}) implies that the  thermodynamic driving force, i.e. the concentration gradient, of any species acts equally on flow of both species. 
When the chemical potentials of both species have the same sign, i.e. the concentration gradients are parallel, they add synergistic to a greater driving force (Eq.~\eqref{freeenergycs}). In Appendix B we show, that an increase of the chemical potential of either species by raising its higher concentration always increases its flow on the CS, which is identical with flow of the other species (Eq.~(\ref{toflowcs}). Hence we have a perfect cooperation. Note that this is not as trivial as it may appear at a first glance. Increasing the higher concentration of any species could also imply that blocking reduces flow of the other species. Vice versa holds when the driving forces of the species have opposite signs. Reducing the magnitude of the whole driving force $|\Delta U|$ reduces flow of either species, until both cease for vanishing $\Delta U$.        

The way towards this extreme coupling of transport of both species and hence, their capability to cooperate is shown in Fig.~(\ref{Phasediagram}) for parallel concentration gradient. We define cooperation when flow of either species mutually profits from an existing parallel directed concentration gradient of the other, i.e.
\begin{eqnarray}
J^{(A)}(\Delta\mu^{(A)},\Delta\mu^{(B)})&>&J^{(A)}(\Delta\mu^{(A)},0)\cr
J^{(B)}(\Delta\mu^{(A)},\Delta\mu^{(B)})&>&J^{(B)}(0,\Delta\mu^{(B)})\;.\cr
\end{eqnarray}  
Vice versa competition on cost of both implies that flow of either species decreases in the presence of a parallel directed gradient of the other species, when compared to a vanishing gradient.
Promotion of one species on cost of the other , e.g. $B$ on cost of $A$, is given, when the parallel directed gradient of the latter induces a higher flow of $B$, compared to a vanishing gradient, however, flow of $A$ is reduced by the gradient of $B$,
\begin{align}
\begin{aligned}
J^{(B)}(\Delta\mu^{(A)},\Delta\mu^{(B)})&>J^{(B)}(0,\Delta\mu^{(B)})\cr
J^{(A)}(\Delta\mu^{(A)},\Delta\mu^{(B)})&<J^{(A)}(\Delta\mu^{(A)},0)\;.\cr
\end{aligned}
\end{align}  
Vice versa turn the greater-than signs when $A$ promoted on cost of $B$.  The Fig.~(\ref{Phasediagram}) shows that with increasing entanglement  of the transport pathways of the two species ($\Delta E$ increases) the range of concentration gradients for which both cooperates increases (pink) whereas those of lossy competition (turquoise) decreases. In the limiting case ($\Delta E \to \infty$) there would be sole cooperation as state space is reduced to CS.     

The dependence of cooperation on the degree of entanglement of species transport holds also in the other direction when opposing concentration gradients force the particles to move in opposite directions. In the extreme case, when there is a perfect coupling of transport in the CS, flows of both vanish for opposing but in magnitude equal concentration gradients ($\Delta\mu^{(A)}=-\Delta\mu^{(B)}$), as there is no net driving force left. This become evident from Fig.~(\ref{Opposingrad}), where the gradient of species $A$ is held constant, and that of $B$ in opposing direction increases. For a vanishing gradient of $B$, the gradient of   $A$ drives $B$ parallel to its direction ($J^{(B)}>0$). Flow of $B$ is positively related to the coupling strength. Increasing an opposing gradient of $B$ ($\Delta\mu^{(B)}=\log(c_1^{(B)}/c_2^{(B)})<0$) monotonically decreases flow of $B$ til cessation,  and change of its sign parallel to that of the gradient. Flow $A$ also decreases. The stronger the coupling, the higher must be the magnitude of the opposing gradient of $B$ to achieve cessation of its flow. For for strong coupling, $\Delta E=20$,  the flow curves of species $A$  and $B$ become almost identical, and both flows cease for equal opposing concentration gradients $c_1^{(B)}/c_2^{(B)}=c_2^{(A)}/c_1^{(A)}=10^{-2}$. 

\section{Entropy production}
\begin{figure*}
\includegraphics[trim = 1mm 1mm 1mm 1mm, clip,width=16cm]{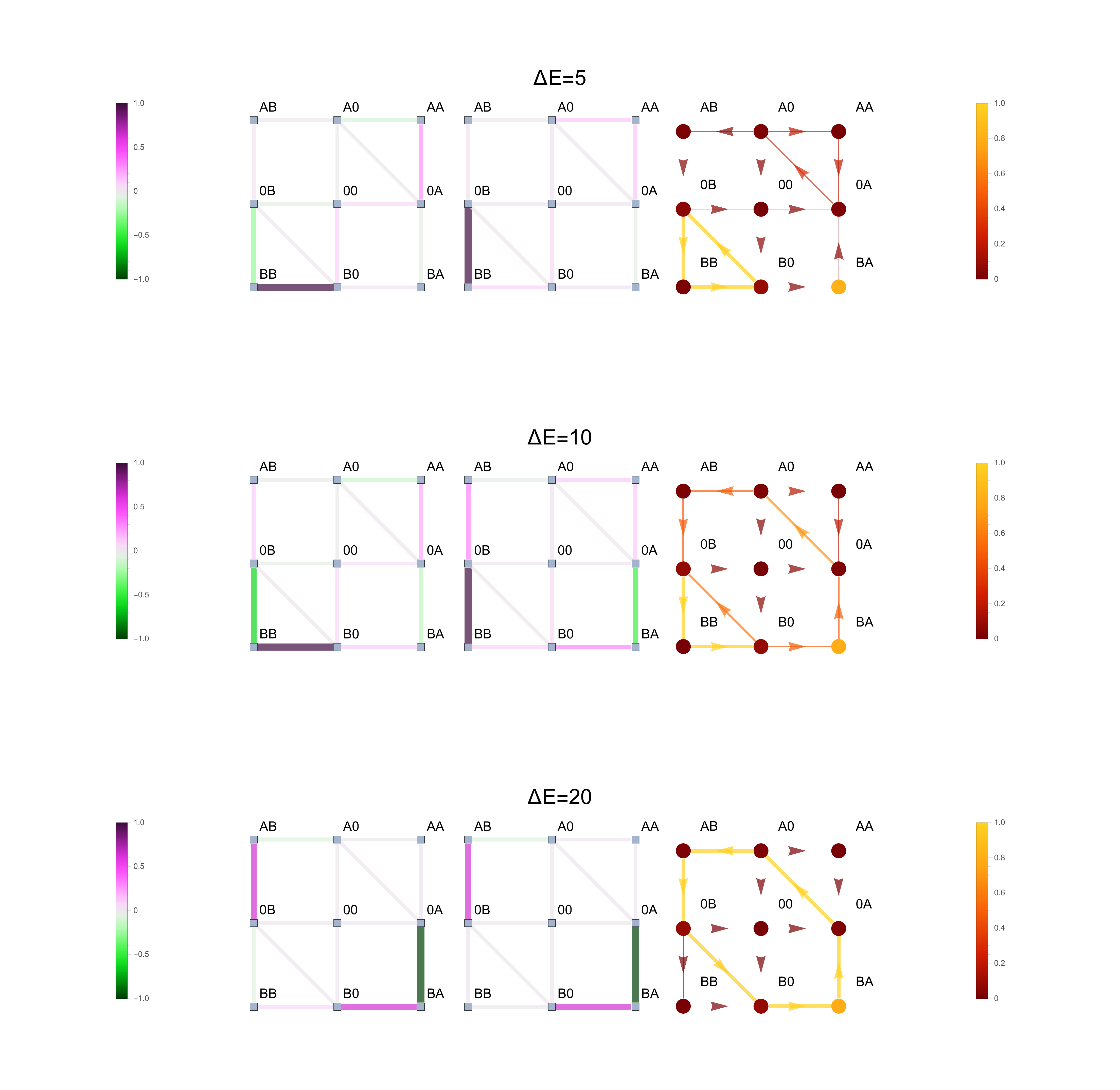}
\caption{Entropy production (left and mid panel) and flow with occupation probability (right panel) within state space as a function of the coupling parameter $\Delta E$. Entropy production is differentiated into overall (left panel) and that solely related to particle exchange (mid panel). The values are normalized to its  maximum magnitude ($\to \dot{S}_{\boldsymbol{\sigma},\varsigma}/\hbox{Max}(|\dot S_{\boldsymbol{\sigma},\varsigma}|)$ and color coded from -1 to +1. Coding of flow and occupation probability see Fig.~(\ref{Flowstsp}). Concentration gradients of $A$ and $B$ are opposing, $c_1^{(A)}/c_2^{(A)}=100, \;c_2^{(B)}/c_1^{(B)}=50$, with $k_+^{(A)} c_2^{(A)}=0.1,\; k_-^{(A)}=1$ and $k_+^{(B)} c_1^{(B)}=e^1\times 0.1,\; k_-^{(B)}=e^{-1}\times 1$, the latter implying a moderate attractive particle-channel interaction $\Phi^{(B)}=-\ln(k_+^{(B)}/k_-^{(B)})=-2$. Note that for loose coupling flows of both species between the baths follow the direction of their gradient,i.e. they are opposing, $(0A)\to (A0)$ vs. $(0B)\leftarrow (B0)$. For high coupling the transition dynamics of the system is restricted to the CS \eqref{limitcycle} and directions of flows are parallel $(0A)\to (A0)$ and  $(0B)\to(B0)$.} 
\label{LocalentropyProd}
\end{figure*}

\begin{figure}
\includegraphics[width=8cm]{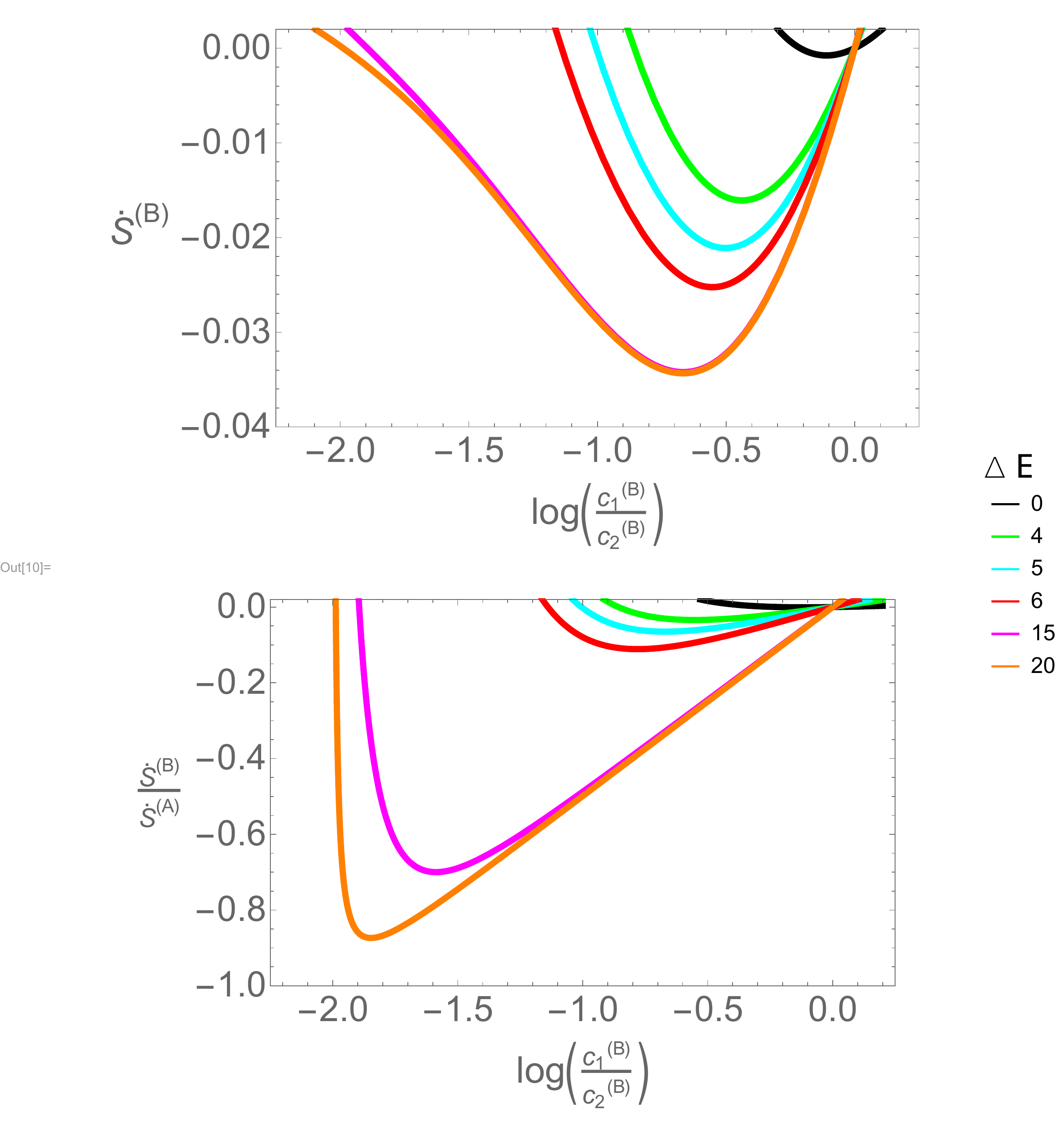}
\caption{Overall negative entropy production of species $B$, $\dot{S}^{(B)}=\Delta\mu^{(B)}J^{(B)}$ (above) related to antiparallel transport against its concentration gradient, which is induced by the driving of species $A$. Below is given the corresponding efficiency quantified by the negative entropy production of $B$ per entropy production related to transport of $A$.  The concentration gradient of $A$ is directed from bath~1 to bath~2,  $c_1^{(A)}/c_2^{(A)}=100$. Particle concentration of $B$ in bath~1 is $k_+^{(B)} c_1^{(B)}=0.1$ and that in bath~2 is elevated about this level. Other parameters of channel dynamics at channel ends of both species as in Fig.~(\ref{LocalentropyProd}). Various coupling strengths $\Delta E$ are considered.For maximum coupling strength the efficiency curve approaches $\dot{S}^{(B)}/\dot{S}^{(A)}\approx\Delta\mu^{(B)}/\Delta\mu^{(A)}=\log(c_1^{(B)}/c_2^{(B)})/\log(100)$ in Eq.~(\ref{maxeff}). } 
\label{EntropyEfficiency}
\end{figure}

To evaluate the thermodynamic coupling of two species transport the entropy production related to state space transitions will be determined. In general entropy production of a system, which is coupled by heat or particle exchange to baths, consists of entropy production within the state space $\boldsymbol{\Sigma}$, which can be measured by changes of the Shannon entropy 
\begin{equation}
S_{\boldsymbol{\Sigma}}=\sum_{\boldsymbol{\sigma}\in\boldsymbol{\Sigma}} -\ln(P_{\boldsymbol{\sigma}})P_{\boldsymbol{\sigma}}\;,\label{shannon}
\end{equation}
with probability $P_{\boldsymbol{\sigma}}$ to find the system in the state $\boldsymbol{\sigma}$, and the entropy production, mediated by the system, within in the baths, 
\begin{equation}
\dot{S}=\dot{S}_{\boldsymbol{\Sigma}}+\dot{S}_{\hbox{bath}}\;.
\end{equation}
The latter refers to Schnackenberg's entropy production \cite{Schnackenberg}. For a transition between two states $\boldsymbol{\sigma}\leftarrow\boldsymbol{\varsigma} $ it is determined by the corresponding flow $J_{\boldsymbol{\sigma},\boldsymbol{\varsigma}}$ and the free energy difference $\Delta\epsilon_{\boldsymbol{\sigma},\boldsymbol{\varsigma}}$ (see Eqs.~(\ref{freenergydiff},\ref{flows})) between the states in direction of the flow,
 \begin{equation}
 \dot{S}_{\boldsymbol{\sigma},\boldsymbol{\varsigma}}=-\Delta\epsilon_{\boldsymbol{\sigma},\boldsymbol{\varsigma}}J_{\boldsymbol{\sigma},\boldsymbol{\varsigma}}\;.\label{entropyproductionSP}
\end{equation}    
The latter equation is easily understood, e.g. when $\Delta\epsilon$ is an energetic difference, Eq.~(\ref{entropyproductionSP}) describes the heat production per time in the bath. When $\Delta\epsilon$ is related to particle exchange, it is 
a step within the mixing entropy process. Note that a prerequisite for application of Eq.~\eqref{entropyproductionSP} is the assumption of instantaneous equilibration of heat or particle concentrations within the baths.
As we consider steady state conditions, the Shannon entropy within state space in Eq.~(\ref{shannon}) remains constant. Hence, the whole entropy production reduces to the sum over the local ones within the baths, i.e.
\begin{equation}
\dot{S}=\frac{1}{2}\sum_{\boldsymbol{\sigma},\boldsymbol{\varsigma}\in \boldsymbol{\Sigma}} \dot{S}_{\boldsymbol{\sigma},\boldsymbol{\varsigma}}\;.\label{entropyproduction2}
\end{equation}
The factor $1/2$ derives from the fact that $\Delta\epsilon_{\boldsymbol{\sigma},\boldsymbol{\varsigma}}$, and $J_{\boldsymbol{\sigma},\boldsymbol{\varsigma}}$ change concordant signs, when states $\boldsymbol{\sigma}$ and $\boldsymbol{\varsigma}$ are interchanged, i.e. $\dot{S}_{\boldsymbol{\sigma},\boldsymbol{\varsigma}}=\dot{S}_{\boldsymbol{\varsigma},\boldsymbol{\sigma}}$ . 
For the further evaluation of the sum, we separately consider transitions which are involved in particle exchange between bath~1 and the adjacent channel end, i.e. for species $X$, $(s,0)\rightleftharpoons (s,X)$, with $s=A,\;B,\;0$. The corresponding free energy difference (Eq.~(\ref{freenergydiff})) may be written as
\begin{eqnarray}
\Delta\epsilon_{\boldsymbol{\sigma},\boldsymbol{\varsigma}}&=&-\sign(\boldsymbol{\sigma},\boldsymbol{\varsigma})\ln\left(\frac{k_+ c_1^{(X})}{k_-}\right)\cr\cr &=&\sign(\boldsymbol{\sigma},\boldsymbol{\varsigma})\left(-\ln\left(\frac{k_+ c_2^{(X})}{k_-}\right)-\ln\left(\frac{c_1^{(X}}{c_2^{(X}}\right)\right)\cr\cr
&=&\Delta\epsilon_{\boldsymbol{\sigma},\boldsymbol{\varsigma}}^{(e)}-\Delta\mu^{(X)}\sign(\boldsymbol{\sigma},\boldsymbol{\varsigma})\;. \label{freeenergydifference}
\end{eqnarray}
The function ``$\sign$'' adjusts the sign of the free energy difference, which is 1 when particles enter the channel from bath~1, $(s,X)\leftarrow (s,0)$ and vice versa $-1$ for $(s,X)\to (s,0)$ transitions. 
According to Eq.~\eqref{freeenergydifference} the free energy difference between states $\boldsymbol{\sigma},\boldsymbol{\varsigma}$ is that that would be present under equilibrium conditions, $\Delta\epsilon_{\boldsymbol{\sigma},\boldsymbol{\varsigma}}^{(e)}$, i.e. when concentrations in both baths equal $c_2^{(X)}$, minus the difference of chemical potentials between both baths. All other transitions, and hence, corresponding free energy differences, between states which are not involved in particle exchange with bath~1 do not differ from their value under equilibrium conditions. Under equilibrium conditions the states may be assigned a potential $\phi_{\boldsymbol{\sigma}}$  (Eq.~(\ref{equilibrium})), the differences of which determines the free energy difference between the states (Eq.~(\ref{potential})).  Free energy differences between states involved in particle transport with bath~1 have to be adjusted by the difference of the chemical potential. Inserting this into Eq.~\eqref{entropyproductionSP} enables to rewrite the entropy production in Eq.~(\ref{entropyproduction2})as 
\begin{eqnarray}
\dot{S} &=&  \frac{1}{2}\sum_{\hbox{\tiny{with bath~1}}}(-\Delta\epsilon_{\boldsymbol{\sigma},\boldsymbol{\varsigma}}^{(e)}+\sign(\boldsymbol{\sigma},\boldsymbol{\varsigma})\Delta\mu^{(X)})\;J_{\boldsymbol{\sigma},\boldsymbol{\varsigma}}+\cr 
&&\frac{1}{2}\sum_{\hbox{\tiny{not with bath~1}}}-\Delta\epsilon_{\boldsymbol{\sigma},\boldsymbol{\varsigma}}^{(e)}\;J_{\boldsymbol{\sigma},\boldsymbol{\varsigma}}\cr\cr
&=& \frac{1}{2}\sum_{\hbox{\tiny{with bath~1}}}\Delta\mu^{(X)}\sign(\boldsymbol{\sigma},\boldsymbol{\varsigma})\;J_{\boldsymbol{\sigma},\boldsymbol{\varsigma}}+\frac{1}{2}\cr
&&\left(\sum_{\hbox{\tiny{with bath~1}}}(-\phi_{\boldsymbol{\sigma}}+\phi_{\boldsymbol{\varsigma}})\;J_{\boldsymbol{\sigma},\boldsymbol{\varsigma}}+\sum_{\hbox{\tiny{not with bath~1}}}(-\phi_{\boldsymbol{\sigma}}+\phi_{\boldsymbol{\varsigma}})\;J_{\boldsymbol{\sigma},\boldsymbol{\varsigma}}\right)\cr\cr
&=&\frac{1}{2}\sum_{\hbox{\tiny{with bath~1}}}\Delta\mu^{(X)}\sign(\boldsymbol{\sigma},\boldsymbol{\varsigma})\;J_{\boldsymbol{\sigma},\boldsymbol{\varsigma}}+\cr
&&\frac{1}{2}\sum_{\boldsymbol{\sigma},\boldsymbol{\varsigma}\in \boldsymbol{\Sigma}}(-\phi_{\boldsymbol{\sigma}}+\phi_{\boldsymbol{\varsigma}})\;J_{\boldsymbol{\sigma},\boldsymbol{\varsigma}}\;.\cr\label{entropysum}
\end{eqnarray}
 The latter terms vanishes as in the steady state flow is conserved around a state (Kirchoff's circuit rule, see Eq.~\eqref{Kirchoff}), i.e.  
 \begin{equation}
 \sum_{\boldsymbol{\sigma},\boldsymbol{\varsigma}\in \boldsymbol{\Sigma}}\phi_{\boldsymbol{\sigma}}\;J_{\boldsymbol{\sigma},\boldsymbol{\varsigma}}=\sum_{\boldsymbol{\sigma}\in \boldsymbol{\Sigma}}\phi_{\boldsymbol{\sigma}}\underbrace{\sum_{\boldsymbol{\varsigma}\in \boldsymbol{\Sigma}}\;J_{\boldsymbol{\sigma},\boldsymbol{\varsigma}}}_{=0}=0\;.
 \end{equation}
Note that the same is true for $\sum_{\boldsymbol{\sigma},\boldsymbol{\varsigma}\in \boldsymbol{\Sigma}}\phi_{\boldsymbol{\varsigma}}\;J_{\boldsymbol{\sigma},\boldsymbol{\varsigma}}$, as $J_{\boldsymbol{\sigma},\boldsymbol{\varsigma}}=-J_{\boldsymbol{\varsigma},\boldsymbol{\sigma}}$.
The flows in the first term of Eq.~(\ref{entropysum}) consist of those for species $A$ and $B$, 
\begin{eqnarray}
\frac{1}{2}\sum_{\hbox{\tiny{with bath~1}}}\Delta\mu^{(X)}\;J_{\boldsymbol{\sigma},\boldsymbol{\varsigma}}&=&\Delta\mu^{(A)} \frac{1}{2}\sum_{A \hbox{\tiny{with bath~1}}} \sign(\boldsymbol{\sigma},\boldsymbol{\varsigma}) J_{\boldsymbol{\sigma},\boldsymbol{\varsigma}}+\cr
&&\Delta\mu^{(B)} \frac{1}{2}\sum_{B \hbox{\tiny{with bath~1}}} \sign(\boldsymbol{\sigma},\boldsymbol{\varsigma})J_{\boldsymbol{\sigma},\boldsymbol{\varsigma}}\nonumber
\end{eqnarray}
The sum over the particle exchange flows at the channel ends is just twice the steady state flow $J^{(A)},\; J^{(B)}$ of the respective species between the baths (see Eq.~\eqref{Pflow}).  Finally we get for the entropy production  
\begin{equation}
\dot{S}=\Delta\mu^{(A)} J^{(A)}+\Delta\mu^{(B)} J^{(B)}\;.\label{mixingentropy}
\end{equation}
This equation states that the sum over the particular entropy productions in state space is that of the mixing entropy production in the baths. Conversely this equation states that the sources of the overall mixing entropy production may be allocated to those in state space. 

In Fig.~(\ref{LocalentropyProd}) opposing gradients determine particle transport, from bath~1 to baht~2 for species $A$ and vv for species $B$. For a moderate coupling of the transport pathways of the two species  ($\Delta E=5$) respective flows are parallel with their gradient, $(0A)\to (A0)$ and $(0B)\leftarrow (B0)$. 
The loose coupling enables that each species follows their thermodynamic driving force rather unmolested by the other. The preferred closed paths in state-space which are involved in particle transport between the baths are $(0A)\to(A0)\to (AA){\color{gray}{\to (0A)\cdots}} $ for species $A$, and ${\color{gray}{\cdots (0B)\leftarrow}} (B0)\leftarrow (BB)\leftarrow (0B)$ for species $B$, respectively. Note that the higher flow of species $B$ when compared to that of $A$ is induced by its attractive particle channel interaction $\Delta\Phi^{(B)}=-\ln (k_+^{(B)}/k_-^{(B)})=-2$. Closed paths including the empty channel state $(00)$ are less frequented. Entropy production is negative for transitions from baths with higher particle concentrations towards states occupied by two particles of the same species. i.e. $(0B)\to(BB)$, and $(A0)\to(AA)$. This is due to the fact that flow is directed to theses states with a higher energetic level, $\Delta E=5$ for $A$ and $\Delta E-\Delta\Phi^{(B)}=5-2=3 $ for $B$, which cannot be compensated by the fraction of free energy difference related to particle exchange. The purely energetic component of negative entropy production is exactly balanced by a positive entropy production when these states set particles free towards the baths in direction of their gradient, $(BB)\to (B0)$ and $(AA)\to(0A)$. The part of entropy production solely related to particle exchange, is, in contrast, always positive around the states occupied by two like particles.  

With increasing coupling ($\Delta E=10$) flow towards states occupied by two unlike particles are favored, on cost of flow towards states with two like particles. The entropy production is similar to the situation for $\Delta E= 5$ except that there is a negative particle exchange related entropy production for species $B$ involved transitions ($(A0)\to (AB)$) and ($(BA)\to (0A)$). Responsible is the increasing flow between theses states which is anti parallel directed to the thermodynamic driving force, namely the free energy differences related to particle exchange.  These flows contribute to a fraction of particle flow against the concentration gradient of $B$, though its overall flow is still parallel to it  $(0B)\leftarrow (B0)$.

With a high coupling ($\Delta E=10$), flow is almost solely present on the cyclic state space CS (\ref{limitcycle}),where it is constant. As flow towards states occupied by two like particles has almost ceased, entropy production consists solely of its  particle exchange related component. The negative entropy production ($(A0)\to (AB)$, $(BA)\to (0A)$) is now in line with positive entropy production ($(AB)\to (0B)$, $(B0)\to (BA)$). The negative entropy production is related to transport of $B$ against its gradient, the positive entropy production, which acts as the thermodynamic motor, drives $A$ in direction of its concentration gradient. The sum of both must be positive, i.e. the whole entropy production the system generates in the baths (Eq.~ (\ref{mixingentropy})), is positive to satisfy the 2nd law of thermodynamics.

The dependence of the whole entropy production  of $B$ on its gradient which is opposed to that of the driving species $A$ and the coupling parameter $\Delta E$ is shown in Fig.~(\ref{EntropyEfficiency}). Negative entropy production is present when gradient and flow of $B$ are anti parallel, the latter due to the driving of $A$. Entropy production is zero for a vanishing gradient of $B$, $c_2^{(B)}=c_1^{(B)}$. With increasing opposed gradient of $B$, entropy production reaches a minimum (or negative entropy production a maximum) and vanishes when this gradient is strong enough to make flow of $B$ cease, i.e. when it balances the driving effect of $A$. With increasing coupling strength the magnitude of this ``ceasing´´ gradient of $B$ increases until for $\Delta E\to \infty$ it approaches that of $A$, i.e. $|\Delta\mu^{(B)}_{\hbox{cease}}|\to|\Delta\mu^{(A)}|$. The reason is that an increasing coupling strength confines the relevant state space to the CS (\ref{limitcycle}). Within this cyclic state space the free energy gain, $\Delta\mu^{(A)}$, and loss, $\Delta\mu^{(B)}$, are in line. Flows of $B$ and $A$ are identical (Eq.~(\ref{toflowcs})), i.e. both cease when the opposed gradients have the same magnitude, $\Delta\mu^{(B)}=-\Delta\mu^{(A)}$. 
The efficiency of $A$ to drive $B$ against its gradient, i.e. its capability to create negative entropy production for $B$ by pumping it against its gradient on cost of its own positive entropy production $\dot{S}^{(B)}/\dot{S}^{(A)} $, increases with the coupling strength (see Fig.~\eqref{EntropyEfficiency}). For strong coupling it approaches 
\begin{eqnarray}
\eta&=&\lim_{\Delta E\to\infty}\frac{\dot{S}^{(B)}}{\dot{S}^{(A)}}\cr\cr
&=&\frac{\Delta\mu^{(B)}}{\Delta\mu^{(A)}}\;\underbrace{\lim_{\Delta E\to\infty}\;\frac{J^{(B)}}{J^{(A)}}}_{J^{(A)},\;J^{(B)}\to J^{(CS)}}\cr\cr
&=&\frac{\Delta\mu^{(B)}}{\Delta\mu^{(A)}}\;,\label{maxeff}
\end{eqnarray}
as flows of $A$ and $B$ become equivalent in this case (Eq.~(\ref{toflowcs}). 
 
\section{Summary and Discussion}
In a minimalist model of a channel transport of two species between two baths, we investigated how interspecies cooperation and competition depend on the entanglement of the respective transport properties. Transition dynamics between channel states were described as a Markov process in a 9-D state space, which in contrast to mean field approaches conserves interspatial correlations of interparticle interactions. The entanglement of the transport of the species was varied by increasing the affinity of the empty channel to absorb any particle and a repulsive intraspecies interaction, which hampers occupation of the channel by particles of the same species. This procedure favors occupation of the channel by single particles of any species, and  by two particles of different species, which also couples the transport of the two species. With increasing coupling, the transition dynamics is almost confined to a cyclic sub space (CS) within the state space. Here, the perfect coupling of the two species transport causes their particle flows to be equivalent. For parallel concentration gradients of the species it was shown, that toward this limiting case the capability of the system for cooperation increases. This implied that mutually flow of either species increased with increasing gradient of the other species. Conversely is the situation when for opposing concentration gradients the coupling strength increases. Magnitude of flow of the species which is driven by the higher concentration gradient eventually decreases whereas flow of the other species increases till they become equivalent. We considered global and local entropy production, i.e. with respect to transitions in state space. For loose coupling opposing gradients drive their species rather independently from the other through the channel, and entropy production of both is positive. The stronger the coupling the more the species under the influence of the stronger gradient drives the other one anti parallel to its gradient, i.e. its entropy production becomes negative, which can be localized in state space. An increase in the coupling strength is also reflected in the increase of the degree of effectiveness, i.e. of the amount of negative entropy production of the driven species per positive entropy production of the driving on.

Particle transport through (nano) channels, which connect two baths, has been extensively worked on in the past. Focus was laid primarily on the impact of particle in-channel interaction. For particles which do not interact within the channel, flow is proportional to the translocation probability, i.e. the probability that  a particle located at one channel end, leaves it at the other \cite{bezrukov2002,*bezrukov2003b}. E.g. for fast dynamics at the channel ends this conditional probability is proportional to $\langle\exp(\Phi(x))/D(x)\rangle^{-1}$, where $\langle\bullet\rangle$ denotes the spatial average along the channel and $\Phi$ is the interaction potential, or more general free energy, when entropic forces are also included, and $D(x)$ the diffusion coefficient \cite{bauer2005,*Bauer2006, *Bauer2010}. This implies that for non interacting particles flow is independent from spatial permutations of the particle in-channel interaction, e.g. independent from the location of an attractive binding site. Additionally flow monotonically increases for attractive interactions ($\Phi(x)<0$). The situation changes when a particle inside the channel impedes access of those from the baths. Flow is then the product of flow, that would be present in the absence of interparticle interaction, and the probability to find the channel non-occupied. The latter decreases with increasing binding strengh of the channel, i.e. it exhibits an opposed dependence compared with that of the flow in the absence of interparticle interaction. So, there is a tradeoff of binding strength, for which flow reaches a maximum \cite{Bauer2006, Bezrukov2007}. The direction of a concentration gradient asymmetrically affects the probability to find the channel non-occupied. So it is lower for a binding site which is located near the bath with the higher concentration, when compared to its symmetric counterpart located near the bath with the lower concentration. Hence, a binding site located in trans position of the concentration gradient implies a higher flow than that in cis position \cite{Bauer2006, Bezrukov2007}. This flux asymmetry may also be extended from binding sites to entropic traps inside the channel \cite{bezrukov2009}. 

The above model, which allows only one particle to occupy the channel, was applied to study the effect of interparticle interaction for two species transport \cite{Bauer2010}. As interparticle interaction consists solely of blocking particles from the baths to assess the channel, selective transport may only be achieved when the two species differ by transport properties or particle in-channel interactions. E.g. a binding site of one species may favor its transport on cost of the other. However, this selectivity is based on pure competition which means that mutually transport of any species would be higher for a vanishing concentration gradient of the other. 

When interparticle interactions are also feasible  within the channel a variety of new effects appear which were addressed by mean field approches and simulations \cite{Zilman2009PRL, Zilman2010PloSComputBiol, Zilman2010}. We recently developed an exactly solvable Markovian model of two species channel transport, which, in contrast to the aforementiond mean field approaches, explicitly conserves spatial correlations between channel sites \cite{Bauer2013}. The species  under the influence of parallel concentration gradient may cooperate, mutually promote the other species on cost of its own, or completely compete for transport. Which kind of regime is present depends among others on the concentration gradients in the baths and the strength of particle channel-interactions. Competition in the presence of parallel concentration gradients means that flow of each species is lower compared with that in the absence of the other's gradient. This may be easily explained by jamming, following the same arguments as for the channel blocking above. The other regimes demand more sophisticated explanations. Promotion of the other species transport is feasible, as the own concentration gradient implies an asymmetric occupation of its particles and, hence, asymmetric sterical interaction profile within the channel. These asymmetric constraints act as an entropic force which biases flow of the other species in direction of the gradient. If this bias is stronger than blocking, the gradient increases flow of the other species. This effect was also shown by simulations \cite{Zilman2010PloSComputBiol}. Finally, the regime of cooperation means, that the effect of promotion is mutually given for both species. We also found, that a longer channel is more capable for cooperation and promotion as it offers more spatial options for particles of different species to interact \cite{Bauer2013}. 

In this paper now, we directly focused on the in-channel interparticle interaction and its implication for two-species particle flow. This was done in the shortest possible channel, namely one with two occupation sites. The simple structure of state space and transitions within  made it easy to identify a path with optimum coupling of transport, and hence, capability for cooperation in case of parallel concentration gradients. Though the 9D-state space appears simple, its relation with the non-conservative thermodynamic driving forces acting within, still offer many issues which have to be addressed. Besides the cyclic sub state space CS, on which the transition dynamics implies perfect transport coupling of the two species, there are other cyclic paths in which free energy is gained. These ``leak´´ flows in state space demand further evaluation.  Another question is, whether there are, perhaps exotic, transition/interaction patterns in state space, for which far away from equilibrium an increase of the gradient of one species implies a reduction in its flow, i.e. the thermodynamic response acts opposite to its direction of force like a Brownian donkey.   
       
\appendix 
\numberwithin{equation}{section} 
\setcounter{equation}{0}
\section{ Flow in cyclic state space}
In order to derive Eq.~(\ref{flowcs}) we start not with a cyclic state space, but with an open linear one which has $N$ positions, the ends of which (position $1$ and $N$) are adjacent to reservoirs (baths) which we label $0$ and $N+1$, respectively. Only nearest neighbor transitions are allowed, and for the ends, also with the reservoirs. The dynamics is that of an stationary Markov process. Though this model holds for any Markovian transition dynamics between states, it is simpler for our understanding to consider the positions as spatial ones, and the interaction at the ends with the reservoirs as particle exchange processes. The reservoirs serve as constant source with concentrations $P_0,\; P_{N+1}$, as well as absorbers of particles. Hop in/out rates from the reservoirs are $\lambda_{1,0} P_0,\;\lambda_{0,1}$ and $\lambda_{N,N+1} P_{N+1},\;\lambda_{N+1,N}$. Transitions within the state space are given by the rates $\lambda_{i\leftarrow j}=\lambda_{i,j}$. So the dynamics of the probability distribution $\boldsymbol{P}=(P_1,\cdots,P_n)^t$ in this state space is determined by
\begin{equation}
\frac{d}{dt}\;\boldsymbol{P}=\boldsymbol{\lambda}\boldsymbol{P}+\begin{pmatrix} \lambda_{1,0\;} P_0\\0\\
\vdots\\0\\
\lambda_{N,N+1} P_{N+1}
\end{pmatrix}\label{Masterappendix}
\end{equation}   
with 
\begin{equation}
\boldsymbol{\lambda}={\tiny{\begin{pmatrix}
-\lambda_{2,1}-\lambda_{0,1} &\lambda_{1,2} &\cdots  &0\\
\lambda_{2,1}& -\lambda_{1,2}-\lambda_{3,2} &\cdots   & 0\\
0 &\lambda_{3,2} &\cdots  &0  \\
0 & 0 & \vdots  &\lambda_{N-1,N}\\
0 & 0   &\cdots  & -\lambda_{N-1,N}-\lambda_{N+1,N}\\
\end{pmatrix}}}\,.
\label{TMatrix1D}
\end{equation}
This transition matrix conserves probability to find a particle within the channel, $\sum_{i}\lambda_{i,j}=0$, except at the ends where transitions to the reservoirs are present. The transition rates define a free energy difference between respective states $\epsilon_{i+1,i}=-\ln(\lambda_{i+1,i}/\lambda_{i,i+1})$. In the open linear topology of state space they can be derived from a potential,  $\epsilon_{i+1,i}=\varphi_{i+1}-\varphi_i$, with $\varphi_j=\sum_{\nu=1}^{j} \epsilon_{\nu,\nu-1}+\varphi_0$. The potential of reservoir `` 0  '', $\varphi_0$ may be set arbitrary . This implies a potential difference between the reservoirs 
\begin{equation}
\Delta U=\varphi_{N+1}-\varphi_0=\sum_{\nu=1}^{N+1} \epsilon_{\nu,\nu-1}=-\ln\left(\prod_{\nu=1}^{N+1}\frac{\lambda_{\nu,\nu-1}}{\lambda_{\nu-1,\nu}}\right)\;.\label{DeltaU}
\end{equation} 
Flow between neighboring states $J_{i+1\leftarrow i}=J_{i+1,i}$ is given by 
\begin{equation}
J_{i+1,i}=\lambda_{i+1,i}P_i-\lambda_{i,i+1}P_{i+1}=0\;.
\end{equation} 
 It is convenient to rewrite the transition rates between neighboring states in terms of potentials,  
 \begin{eqnarray}
 \lambda_{i+1,i}&=&\bar{\lambda}_{i+1} e^{-(\varphi_{i+1}-\varphi_i)/2}\nonumber\\
 \lambda_{i,i+1}&=&\bar{\lambda}_{i+1} e^{-(\varphi_{i}-\varphi_{i+1})/2}\;\hbox{with}\nonumber\\ 
 \bar{\lambda}_{i+1}&=&\sqrt{\lambda_{i+1,i}\lambda_{i,i+1}}
 \end{eqnarray}
 as a measure of mobility between the states. This enables us to write flow between states in terms of activities $a_i$ and 
resistances $R_i$. With $\bar{\varphi}_i=1/2(\varphi_{i+1}+\varphi_i)$ as the mean potential of two nearby states we get 
 \begin{equation}
\underbrace{e^{\varphi_i}P_i}_{=a_i}-\underbrace{e^{\varphi_{i+1}}P_{i+1}}_{=a_{i+1}}= J_{i+1,i} \underbrace{\frac{e^{\bar{\varphi}_i}}{\bar{\lambda}_i}}_{=R_i}\;.\label{activity}
 \end{equation}
In the steady state flow is constant throughout, $J_{i+1,i}\equiv J$. Just by adding up the activity differences in Eq.~(\ref{activity}) flow turns out in form of an Ohm's law, 
\begin{equation}
a_0-a_{N+1}=J\sum_{i=1}^{N+1} R_i\;.\label{ohm}
\end{equation}
However a cyclic state space demands a more sophisticated approach, since, as we will see, the baths will be integrated into state space. It was recently shown that for unidirectional transport, i.e. concentration in one bath vanishes (either $P_{N+1}=0$, or $P_0=0$), the steady state flow fulfills \cite{Hardt}
\begin{eqnarray}
J_{0\to N+1}&=&\mathcal{N}_{1\to N+1}/\tau_{1\to N+1}\\
J_{0\leftarrow N+1}&=&-\mathcal{N}_{0\leftarrow N}/\tau_{0\leftarrow N}
\end{eqnarray}
with $\tau_{1\to N+1}$ as the mean first passage time of a particle which starts at positions $i=1$, is reflected when trying to jump back to bath~$0$, and which is completely absorbed in bath~$N+1 $. Vice versa holds for $\tau_{0\leftarrow N}$. It is noteworthy that this relation describes a situation in which both ends of state space are in exchange with the reservoirs, however the mean first passage times derive from a set up with reflective boundary conditions.  $\mathcal{N}$ is the number of particles within the state space, i.e. $\mathcal{N}=\sum_{i=1}^{N} P_i^{(s)}$ for either direction.   These $P_i^{(s)}$ in the steady state are determined from Eq.~(\ref{Masterappendix}) by setting $\dot{\boldsymbol{P}}=0$, i.e. 
\begin{align}
\boldsymbol{P}_{0\to N+1}^{(s)}&=\boldsymbol{\lambda}^{-1}\;\begin{pmatrix} \lambda_{1,0\;} P_0\\0\\
\vdots\\0\end{pmatrix}\;\hbox{or}\cr\cr \boldsymbol{P}_{0\leftarrow N+1}^{(s)}&=\boldsymbol{\lambda}^{-1}\;\begin{pmatrix} 0\\
\vdots\\0\\ \lambda_{N,N+1} P_{N+1}\end{pmatrix}\;.
\end{align}
Hence, these particle numbers are proportional to the respective concentrations in the reservoirs, $\mathcal{N}_{1\to N+1}\sim P_0$ and $\mathcal{N}_{0\to N}\sim P_{N+1}$. This suggests to introduce specific particle numbers $n$ independent from the bath activities by normalizing the particle number by the respective activities \cite{bauer2005, Bauer2006, Bauer2010}, i.e. 
with 
\begin{eqnarray}
n_{1\to N+1}&=&\frac{\mathcal{N}_{1\to N+1}}{e^{\varphi_0}P_0}\;,\cr\cr
n_{0\leftarrow N}&=&\frac{\mathcal{N}_{0\leftarrow N}}{e^{\varphi_{N+1}}P_{N+1}}\label{sppnumber}
\end{eqnarray}
we get
\begin{eqnarray}
J_{0\to N+1}&=& \phantom{-} a_{0\phantom{+1}}\;n_{1\to N+1}/\tau_{1\to N+1} \\
J_{0\leftarrow N+1}&=&-a_{N+1} n_{0\leftarrow N\phantom{+1}}/\tau_{0\leftarrow N\phantom{+1}}\;.
\end{eqnarray}
For arbitrary concentrations in the baths, steady state flow is the superposition of the two unidirectional flows above. In particular, as flow vanishes for equal activities in the baths, we obtain 
\begin{equation}
\frac{n_{1\to N+1}}{\tau_{1\to N+1}}=\frac{n_{0\leftarrow N}}{\tau_{0\leftarrow N}}=\frac{n}{\tau}\;,\label{Aconduct}
\end{equation} 
where $n=1/2 (n_{1\to N+1}+n_{0\leftarrow N})$ and $\tau=1/2(\tau_{1\to N+1}+\tau_{0\leftarrow N})$ are the symmetrical specific particle number and first passage time. So steady state flow for arbitrary concentrations in the baths takes the form
\begin{equation}
J=\frac{n}{\tau}(a_0-a_{N+1})\label{flowlaw}
\end{equation}
Note that with Eqs.~(\ref{activity},\ref{ohm})
$n/\tau$ is the conductivity (inverse of resistance $R$)
\begin{equation}
n/\tau=R^{-1}=\left(\sum_{i=1}^{N} R_i\right)^{-1}=\left(\sum_{i=1}^{N}\frac{e^{\bar{\varphi}_i}}{\bar{\lambda}_i}\right)^{-1}\label{Aconduct2}
\end{equation}   
The explicit determination of the mean first passage times is a bit tedious and can be looked up in Ref.~(\cite{Nadler85, *Nadler86}). In short: one modifies the transition matrix in Eq.~(\ref{TMatrix1D}), $\boldsymbol{\lambda}\to \boldsymbol{\lambda}'$, so that there are either reflective boundary conditions towards bath~0 for determination of $\tau_{1\to N+1}$ and v.v. towards bath~$N+1$ when $\tau_{0\leftarrow N}$ is considered. This is accomplished by just taking out the appropriate hopping out rates. For determination of $\tau_{1\to N+1}$ one positions a particle at $i=1$ and lets the system evolve, i.e. $\boldsymbol{P}(t)=\exp(\boldsymbol{\lambda}'t)\boldsymbol{e}_1$, with $\boldsymbol{e}_1=(1,0,\cdots,0)^t$. The mean first passage time is defined as the mean time the particle needs to get absorbed in bath~$N+1$. So $\tau_{1\to N+1}=\int_0^{\infty} dt\; t\; (-d p(t)/dt)=\int_0^{\infty} dt\; p(t)$, where $p(t)=(1,1\cdots,1)\boldsymbol{P}(t)=\sum_{i=1}^{N}P_i(t)$ is the probability to find the particle still in state space, i.e. $-dp(t)/dt$ is the fraction absorbed at $t$ by bath~$N+1$. So one gets  
\begin{equation}
\tau_{1\to N+1}=(1,1\cdots,1)\frac{1}{\boldsymbol{\lambda}'}\boldsymbol{e}_1=\boldsymbol{e}_1^t \frac{1}{(\boldsymbol{\lambda}')^{t}}
\begin{pmatrix} 1\\
1\\
\vdots\\
1
\end{pmatrix}
\end{equation}  
The same holds for $\tau_{0\leftarrow N}$. The above matrix equation may be solved directly and we get for the mean first passage times 
\begin{eqnarray}
\tau_{1\to N+1}&=&\sum_{i=1}^{N+1}\sum_{\nu=1}^{N} e^{-\varphi_\nu} \frac{e^{\bar{\varphi}_i}}{\bar{\lambda}_i}\;\theta(i-\nu-1)\\
\tau_{0\leftarrow N}&=&\sum_{i=1}^{N}\sum_{\nu=1}^{N} e^{-\varphi_\nu} \frac{e^{\bar{\varphi}_i}}{\bar{\lambda}_i}\;\theta(\nu-i)\\
\tau&=&\frac{1}{2} \left(\sum_{\nu=1}^{N}e^{-\varphi_\nu}\right)\left(\sum_{i=1}^{N+1}  \frac{e^{\bar{\varphi}_i}}{\bar{\lambda}_i}\right)\label{fptimes}
\end{eqnarray}
where $\theta(x)$ is the unit step function, with $\theta=0$, for $x<0$ and otherwise $\theta=1$.
Note that the above equations are in integral form for continuous diffusion-reaction processes well known (e.g. \cite{Nadler85}) and look similar. However, results cannot be transferred readily to the discrete case as details as the distinction between  the potential $\varphi_i$ of a state $i$ and the mean potential between two states $\bar{\varphi}_i$ appear in the discrete case, whereas both become equivalent in the continuum limit.   

Now we have the tools to tackle the cyclic state space. In the steady state we can modify the model of a linear state space between two reservoirs to a cyclic state space by adding one position, namely that of bath~$0$ and closing the bath positions $0$ and $N+1$ together so that we end up with a ring like state space with $N+1$ positions $0,\;1\cdots N$. This implies that $P_0=P_{N+1}$. One round in this cyclic space implies that the free energy change is $\Delta U$ (see Eq.~(\ref{DeltaU})). Note that the ring topology impedes to define a unique potential from free energy differences, as continuation implies $\varphi_{N+1}=\varphi_0+\Delta U\neq\varphi_0$.  As the off-set potential may be chosen arbitrarily, we set it equal to zero, i.e. $\varphi_0=0$. So, with Eq.~(\ref{flowlaw}) flow in this cyclic space becomes 
\begin{equation}
J=\frac{n}{\tau} P_0 (1-e^{\Delta U})\label{ACSflow}
\end{equation}
In contrast to the open situation between two baths, we now have a closed system, and conservation of probability holds, $\sum_{i=0}^{N}P_i=1$. According to Eq.~(\ref{sppnumber}) we get the constraint
\begin{eqnarray}
1&\overset{!}{=}&P_0+\mathcal{N}_{1\to N+1}+\mathcal{N}_{0\leftarrow N}\cr\cr
&=&P_0 (1+n_{1\to N+1}+n_{0\leftarrow N} e^{\Delta U})
\end{eqnarray}   
Insertion into Eq.~(\ref{ACSflow}) and considering Eq.~(\ref{Aconduct}) then gives for the steady state flow in a cyclic space
\begin{equation}
J=\frac{1-e^{\Delta U}}{\tau_{1\to N+1}+ \tau_{0\leftarrow N}\;e^{\Delta U}+\tau/n}\label{ACSflow2}\;.
\end{equation}
Inserting the first passage times from Eqs.~\eqref{fptimes} and the resistance from Eq.~\eqref{Aconduct2}, finally allows to determine flow according to Eq.~\eqref{flowcs}. 
 
\section{Cooperation and competition in cyclic state space}
To investigate the features of the cyclic state space for cooperation and competition, e.g. when the concentration of one species is enhanced, we pick one transition between two states $j\leftarrow j-1$ within the cycle, and increase the rate in counter clockwise direction by a factor $\alpha>1$
\begin{equation}
\lambda_{j,j-1}'=\alpha \lambda_{j,j-1}\;.
\end{equation}
This can, for example, be accomplished by an increase of particle concentration, $k_+c \to k_+ c'$ with $c'>c$, i.e. $\alpha=c'/c$. The backward rate $\lambda_{j-1,j}$ shall remain unchanged. This implies a change in potentials and mobilities. With 
\begin{equation}
\delta\varphi=-\ln(\alpha)<0
\end{equation} we get    
\begin{equation}
\varphi_i'=\begin{cases}
\varphi_i,  &i<j\\
\varphi_i+\delta\varphi, &i\ge j
\end{cases}\;,\nonumber
\end{equation}
in particular the free energy difference after one round in the cyclic state space changes to, 
\begin{equation}
\Delta U'=\varphi'_{N+1}-\varphi'_0=\Delta U+\delta\varphi\nonumber\;.
\end{equation}
Potentials in-between states change to 
\begin{equation}
\bar{\varphi}_i'=\begin{cases}
\bar{\varphi}_i,  &i<j\\
\bar{\varphi}_j+\delta\varphi/2, &i=j\\
\bar{\varphi}_i+\delta\varphi, &i>j
\end{cases}\;,\nonumber
\end{equation} 
and mobilities to
\begin{equation}
\bar{\lambda}_i'=\begin{cases}
\bar{\lambda}_j\;\sqrt{\alpha}=\bar{\lambda}_j e^{\delta\varphi/2},  &i=j\\
\bar{\lambda}_i &\hbox{else}
\end{cases}\;.
\end{equation} 
We will investigate, how this effects first passage times and by this flow in Eq.~(\ref{ACSflow2}). After explicit evaluation of the unit step function $\theta$ we derive from Eqs.~(\ref{fptimes})
\begin{eqnarray}
\tau_{1\to N+1}'&=&\sum_{\substack{i=2\\i\neq j}}^{N+1}\sum_{\nu=1}^{i-1} \frac{e^{\bar{\varphi'}_i}}{\bar{\lambda'}_i} e^{-\varphi'_\nu} +
\underbrace{\frac{e^{\bar{\varphi'}_j}}{\bar{\lambda'}_j}}_{=e^{\delta\varphi} \frac{e^{\bar{\varphi}_j}}{\bar{\lambda}_j}}\sum_{\nu=1}^{j-1} e^{-\varphi'_\nu}\nonumber\\
&=& \sum_{i=2}^{N+1}\sum_{\nu=1}^{i-1} \frac{e^{\bar{\varphi}_i}}{\bar{\lambda}_i} e^{-\varphi_\nu} e^{\delta\varphi\;\theta(j-\nu-1)\theta(i-j)}\;.
\end{eqnarray}
As $ e^{\delta\varphi}<1$, the inequality 
\begin{equation}
e^{\delta\varphi}\tau_{1\to N+1}<\tau_{1\to N+1}'<\tau_{1\to N+1}\label{Ainequal1}
\end{equation}
holds, i.e, the first passage time in direction $1\to N+1$ becomes shorter, however, it is still longer than a lower boundary given by the factor $e^{\delta\varphi}$. Similarly one can show for the first passage time in the opposite direction  
\begin{equation}
\tau_{0\leftarrow N}'=\sum_{i=1}^{N}\sum_{\nu=i}^{N} \frac{e^{\bar{\varphi}_i}}{\bar{\lambda}_i} e^{-\varphi_\nu} e^{-\delta\varphi\;\theta(\nu-j)\theta(j-i-1)}\;.
\end{equation}  
i.e.
\begin{equation}
e^{-\delta\varphi}\tau_{0\leftarrow N}>\tau_{0\leftarrow N}'>\tau_{0\leftarrow N}\;. \label{Ainequal2}
\end{equation}
For the resistance (Eq.~(\ref{Aconduct2})) one gets
\begin{equation}
(\tau/n)'=\sum_{i=1}^{N}\frac{e^{\bar{\varphi}_i}}{\bar{\lambda}_i} e^{\delta\varphi \theta(i-j)}\;,
\end{equation}
i.e.
\begin{equation}
e^{\delta\varphi}(\tau/n)<(\tau/n)'<(\tau/n)\;. \label{Ainequal3}
\end{equation}

Two constellations are now of interest. First we assume that the rate we increase by elevation of the related particle concentration, points parallel to flow direction on the cyclic state space. So flow points in counter clockwise direction, and $\Delta U<0$ which implies   
\begin{eqnarray}
J'&=&\frac{1-e^{\Delta U'}}{\tau'_{1\to N+1}+ \tau'_{0\leftarrow N}\;e^{\Delta U'}+(\tau/n)'}\nonumber\\
&>&\frac{1-e^{\Delta U}}{\tau_{1\to N+1}+ \tau_{0\leftarrow N}\;e^{\Delta U}+(\tau/n)}=J\;.
\end{eqnarray}
This implies that any increase in rate by elevation of the concentration of one partner in direction of flow, elevates flow in cyclic state space. As flows of both species are identical in CS this implies perfect cooperation over the whole range of concentrations. 

Vice versa is the situation when flow on the cyclic state space directs in clockwise orientation, i.e antiparallel to the direction of the rate we increase by elevation of particle concentration. This implies $J<0$, and hence, $\Delta U>0$ (see Eq.\eqref{ACSflow2}). We choose the magnitude of $\delta\varphi$ small enough so that flow $J'$ preserves its negative sign, i.e. $\Delta U'=\Delta U+\delta\varphi  >0$. For the denominator in Eq.(\ref{ACSflow2}) the inequalities (\ref{Ainequal1},\ref{Ainequal2},\ref{Ainequal3}) imply
\begin{equation}
\begin{split}
&\frac{1}{\tau'_{1\to N+1}+ \tau'_{0\leftarrow N}\;e^{\Delta U'}+(\tau/n)'}\\ &< \frac{e^{-\delta\varphi}}{\tau_{1\to N+1}+ \tau_{0\leftarrow N}\;e^{\Delta U}+(\tau/n)}
\end{split}
\end{equation}
Multiplying with $1-e^{U'}$ (note as its sign is negative the greater-than-sign changes changes its direction) leads to
\begin{eqnarray}
J'&=&\frac{1-e^{U'}}{\tau'_{1\to N+1}+ \tau'_{0\leftarrow N}\;e^{\Delta U'}+(\tau/n)'}\nonumber\\
&>&\frac{e^{-\delta\varphi}-e^{\Delta U}}{\tau_{1\to N+1}+ \tau_{0\leftarrow N}\;e^{\Delta U}+(\tau/n)}\nonumber\\
&>&\frac{1-e^{\Delta U}}{\tau_{1\to N+1}+ \tau_{0\leftarrow N}\;e^{\Delta U}+(\tau/n)}=J\;.
\end{eqnarray}
So the anti parallel directed increase of rate increases the negative flow towards zero, i.e. it decreases its magnitude. 
In summary we demonstrated that an isolated increase of rate, in our model by elevation of the appropriate particle concentration, in the cyclic state space implies a higher magnitude of flow when this rate points in direction of flow, and v.v. magnitude of flow is decreased when rate and flow are antiparallel.  

\subsection*{Acknowledgment}
The authors wants to thank his mentor Walter Nadler, Institute for Advanced Simulation (IAS), Juelich Supercomputing Centre (JSC), Forschungszentrum Juelich, D-52725 J\"ulich, Germany, for his always stimulating, imaginative, visionary discussions and support. Walter Nadler deceased June 2015.
The author's research was supported by the Deutsche Forschungsgemeinschaft (SFB688, TBB05 to WB) and the Bundesministerium f\"ur Bildung und Forschung (BMBF01 EO1504 to WB).

 \bibliographystyle{apsrev4-1}
\bibliography{literature}

\begin{thebibliography}{21}%
\makeatletter
\providecommand \@ifxundefined [1]{%
 \@ifx{#1\undefined}
}%
\providecommand \@ifnum [1]{%
 \ifnum #1\expandafter \@firstoftwo
 \else \expandafter \@secondoftwo
 \fi
}%
\providecommand \@ifx [1]{%
 \ifx #1\expandafter \@firstoftwo
 \else \expandafter \@secondoftwo
 \fi
}%
\providecommand \natexlab [1]{#1}%
\providecommand \enquote  [1]{``#1''}%
\providecommand \bibnamefont  [1]{#1}%
\providecommand \bibfnamefont [1]{#1}%
\providecommand \citenamefont [1]{#1}%
\providecommand \href@noop [0]{\@secondoftwo}%
\providecommand \href [0]{\begingroup \@sanitize@url \@href}%
\providecommand \@href[1]{\@@startlink{#1}\@@href}%
\providecommand \@@href[1]{\endgroup#1\@@endlink}%
\providecommand \@sanitize@url [0]{\catcode `\\12\catcode `\$12\catcode
  `\&12\catcode `\#12\catcode `\^12\catcode `\_12\catcode `\%12\relax}%
\providecommand \@@startlink[1]{}%
\providecommand \@@endlink[0]{}%
\providecommand \url  [0]{\begingroup\@sanitize@url \@url }%
\providecommand \@url [1]{\endgroup\@href {#1}{\urlprefix }}%
\providecommand \urlprefix  [0]{URL }%
\providecommand \Eprint [0]{\href }%
\providecommand \doibase [0]{http://dx.doi.org/}%
\providecommand \selectlanguage [0]{\@gobble}%
\providecommand \bibinfo  [0]{\@secondoftwo}%
\providecommand \bibfield  [0]{\@secondoftwo}%
\providecommand \translation [1]{[#1]}%
\providecommand \BibitemOpen [0]{}%
\providecommand \bibitemStop [0]{}%
\providecommand \bibitemNoStop [0]{.\EOS\space}%
\providecommand \EOS [0]{\spacefactor3000\relax}%
\providecommand \BibitemShut  [1]{\csname bibitem#1\endcsname}%
\let\auto@bib@innerbib\@empty
\bibitem [{\citenamefont {Iqbal}\ \emph {et~al.}(2007)\citenamefont {Iqbal},
  \citenamefont {Akin},\ and\ \citenamefont {Bashir}}]{Iqbal}%
  \BibitemOpen
  \bibfield  {author} {\bibinfo {author} {\bibfnamefont {S.~M.}\ \bibnamefont
  {Iqbal}}, \bibinfo {author} {\bibfnamefont {D.}~\bibnamefont {Akin}}, \ and\
  \bibinfo {author} {\bibfnamefont {R.}~\bibnamefont {Bashir}},\ }\href@noop {}
  {\bibfield  {journal} {\bibinfo  {journal} {Nat Nanotechnol}\ }\textbf
  {\bibinfo {volume} {2}},\ \bibinfo {pages} {243} (\bibinfo {year}
  {2007})}\BibitemShut {NoStop}%
\bibitem [{\citenamefont {Jovanovic-Talisman}\ \emph
  {et~al.}(2009)\citenamefont {Jovanovic-Talisman}, \citenamefont
  {Tetenbaum-Novatt}, \citenamefont {McKenney}, \citenamefont {Zilman},
  \citenamefont {Peters}, \citenamefont {Rout},\ and\ \citenamefont
  {Chait}}]{Talisman2009nat}%
  \BibitemOpen
  \bibfield  {author} {\bibinfo {author} {\bibfnamefont {T.}~\bibnamefont
  {Jovanovic-Talisman}}, \bibinfo {author} {\bibfnamefont {J.}~\bibnamefont
  {Tetenbaum-Novatt}}, \bibinfo {author} {\bibfnamefont {A.~S.}\ \bibnamefont
  {McKenney}}, \bibinfo {author} {\bibfnamefont {A.}~\bibnamefont {Zilman}},
  \bibinfo {author} {\bibfnamefont {R.}~\bibnamefont {Peters}}, \bibinfo
  {author} {\bibfnamefont {M.~P.}\ \bibnamefont {Rout}}, \ and\ \bibinfo
  {author} {\bibfnamefont {B.~T.}\ \bibnamefont {Chait}},\ }\href@noop {}
  {\bibfield  {journal} {\bibinfo  {journal} {Nature}\ }\textbf {\bibinfo
  {volume} {457}},\ \bibinfo {pages} {1023} (\bibinfo {year}
  {2009})}\BibitemShut {NoStop}%
\bibitem [{\citenamefont {Bauer}\ and\ \citenamefont
  {Nadler}(2013)}]{Bauer2013}%
  \BibitemOpen
  \bibfield  {author} {\bibinfo {author} {\bibfnamefont {W.~R.}\ \bibnamefont
  {Bauer}}\ and\ \bibinfo {author} {\bibfnamefont {W.}~\bibnamefont {Nadler}},\
  }\href@noop {} {\bibfield  {journal} {\bibinfo  {journal} {Physical Review
  E}\ }\textbf {\bibinfo {volume} {88}},\ \bibinfo {pages} {010703} (\bibinfo
  {year} {2013})}\BibitemShut {NoStop}%
\bibitem [{Note1()}]{Note1}%
  \BibitemOpen
  \bibinfo {note} {We consider the gain of free energy as the negative of the
  corresponding free energy difference $\Delta \epsilon $, i.e. in this case
  $\Delta \epsilon =-\Delta \mu $}\BibitemShut {NoStop}%
\bibitem [{Note2()}]{Note2}%
  \BibitemOpen
  \bibinfo {note} {Note that we normalize all energetic quantities to $kT$,
  i.e. $\Phi \to \Phi /kT$, with $k$ as the Boltzmann constant and $T$ as the
  temperature}\BibitemShut {NoStop}%
\bibitem [{\citenamefont {Nadler}\ and\ \citenamefont
  {Schulten}(1986)}]{Nadler86}%
  \BibitemOpen
  \bibfield  {author} {\bibinfo {author} {\bibfnamefont {W.}~\bibnamefont
  {Nadler}}\ and\ \bibinfo {author} {\bibfnamefont {K.}~\bibnamefont
  {Schulten}},\ }\href@noop {} {\bibfield  {journal} {\bibinfo  {journal} {J
  Chem Phys}\ }\textbf {\bibinfo {volume} {84}},\ \bibinfo {pages} {4015}
  (\bibinfo {year} {1986})}\BibitemShut {NoStop}%
\bibitem [{Note3()}]{Note3}%
  \BibitemOpen
  \bibinfo {note} {We notate the sequence of state variables of a transition
  $\protect \boldsymbol {\sigma }\leftarrow \protect \boldsymbol {\varsigma }$
  as $\protect \boldsymbol {\sigma },\protect \boldsymbol {\varsigma }$ in the
  index of the transition rate $\lambda $ to be in line with the usual index
  notation in matrix algebra}\BibitemShut {NoStop}%
\bibitem [{Note4()}]{Note4}%
  \BibitemOpen
  \bibinfo {note} {For counterclockwise/clockwise direction select the state
  $\protect \boldsymbol {\sigma }_{i+1}$ / $\protect \boldsymbol {\sigma
  }_{i-1}$ as a start and impede transitions from this state in
  clockwise/counterclockwise direction (reflective boundary conditions). $\tau
  _+/-$ is then the mean time to reach $\protect \boldsymbol {\sigma }_i$ for
  the first time (absorptive boundary conditions)}\BibitemShut {NoStop}%
\bibitem [{\citenamefont {Schnackenberg}(1976)}]{Schnackenberg}%
  \BibitemOpen
  \bibfield  {author} {\bibinfo {author} {\bibfnamefont {J.}~\bibnamefont
  {Schnackenberg}},\ }\href@noop {} {\bibfield  {journal} {\bibinfo  {journal}
  {Reviews of Modern Physics}\ }\textbf {\bibinfo {volume} {48}},\ \bibinfo
  {pages} {571} (\bibinfo {year} {1976})}\BibitemShut {NoStop}%
\bibitem [{\citenamefont {Berezhkovskii}\ \emph {et~al.}(2002)\citenamefont
  {Berezhkovskii}, \citenamefont {Pustovoit},\ and\ \citenamefont
  {Bezrukov}}]{bezrukov2002}%
  \BibitemOpen
  \bibfield  {author} {\bibinfo {author} {\bibfnamefont {A.~M.}\ \bibnamefont
  {Berezhkovskii}}, \bibinfo {author} {\bibfnamefont {M.~A.}\ \bibnamefont
  {Pustovoit}}, \ and\ \bibinfo {author} {\bibfnamefont {S.~M.}\ \bibnamefont
  {Bezrukov}},\ }\href@noop {} {\bibfield  {journal} {\bibinfo  {journal} {J
  Chem Phys}\ }\textbf {\bibinfo {volume} {109}},\ \bibinfo {pages} {9952}
  (\bibinfo {year} {2002})}\BibitemShut {NoStop}%
\bibitem [{\citenamefont {Berezhkovskii}\ \emph {et~al.}(2003)\citenamefont
  {Berezhkovskii}, \citenamefont {Pustovoit},\ and\ \citenamefont
  {Bezrukov}}]{bezrukov2003b}%
  \BibitemOpen
  \bibfield  {author} {\bibinfo {author} {\bibfnamefont {A.~M.}\ \bibnamefont
  {Berezhkovskii}}, \bibinfo {author} {\bibfnamefont {M.~A.}\ \bibnamefont
  {Pustovoit}}, \ and\ \bibinfo {author} {\bibfnamefont {S.~M.}\ \bibnamefont
  {Bezrukov}},\ }\href@noop {} {\bibfield  {journal} {\bibinfo  {journal} {J
  Chem Phys}\ }\textbf {\bibinfo {volume} {119}},\ \bibinfo {pages} {3943}
  (\bibinfo {year} {2003})}\BibitemShut {NoStop}%
\bibitem [{\citenamefont {Bauer}\ and\ \citenamefont
  {Nadler}(2005)}]{bauer2005}%
  \BibitemOpen
  \bibfield  {author} {\bibinfo {author} {\bibfnamefont {W.~R.}\ \bibnamefont
  {Bauer}}\ and\ \bibinfo {author} {\bibfnamefont {W.}~\bibnamefont {Nadler}},\
  }\href@noop {} {\bibfield  {journal} {\bibinfo  {journal} {J Chem Phys}\
  }\textbf {\bibinfo {volume} {122}},\ \bibinfo {pages} {244904} (\bibinfo
  {year} {2005})}\BibitemShut {NoStop}%
\bibitem [{\citenamefont {Bauer}\ and\ \citenamefont
  {Nadler}(2006)}]{Bauer2006}%
  \BibitemOpen
  \bibfield  {author} {\bibinfo {author} {\bibfnamefont {W.~R.}\ \bibnamefont
  {Bauer}}\ and\ \bibinfo {author} {\bibfnamefont {W.}~\bibnamefont {Nadler}},\
  }\href@noop {} {\bibfield  {journal} {\bibinfo  {journal} {Proc Natl Acad Sci
  U S A}\ }\textbf {\bibinfo {volume} {103}},\ \bibinfo {pages} {11446}
  (\bibinfo {year} {2006})}\BibitemShut {NoStop}%
\bibitem [{\citenamefont {Bauer}\ and\ \citenamefont
  {Nadler}(2010)}]{Bauer2010}%
  \BibitemOpen
  \bibfield  {author} {\bibinfo {author} {\bibfnamefont {W.~R.}\ \bibnamefont
  {Bauer}}\ and\ \bibinfo {author} {\bibfnamefont {W.}~\bibnamefont {Nadler}},\
  }\href@noop {} {\bibfield  {journal} {\bibinfo  {journal} {PLoSONE}\ }\textbf
  {\bibinfo {volume} {5}},\ \bibinfo {pages} {e15160} (\bibinfo {year}
  {2010})}\BibitemShut {NoStop}%
\bibitem [{\citenamefont {Bezrukov}\ \emph {et~al.}(2007)\citenamefont
  {Bezrukov}, \citenamefont {Berezhkovskii},\ and\ \citenamefont
  {Szabo}}]{Bezrukov2007}%
  \BibitemOpen
  \bibfield  {author} {\bibinfo {author} {\bibfnamefont {S.~M.}\ \bibnamefont
  {Bezrukov}}, \bibinfo {author} {\bibfnamefont {A.~M.}\ \bibnamefont
  {Berezhkovskii}}, \ and\ \bibinfo {author} {\bibfnamefont {A.}~\bibnamefont
  {Szabo}},\ }\href@noop {} {\bibfield  {journal} {\bibinfo  {journal} {J Chem
  Phys}\ }\textbf {\bibinfo {volume} {127}},\ \bibinfo {pages} {115101}
  (\bibinfo {year} {2007})}\BibitemShut {NoStop}%
\bibitem [{\citenamefont {Berezhkovskii}\ \emph {et~al.}(2009)\citenamefont
  {Berezhkovskii}, \citenamefont {Pustovoit},\ and\ \citenamefont
  {Bezrukov}}]{bezrukov2009}%
  \BibitemOpen
  \bibfield  {author} {\bibinfo {author} {\bibfnamefont {A.~M.}\ \bibnamefont
  {Berezhkovskii}}, \bibinfo {author} {\bibfnamefont {M.~A.}\ \bibnamefont
  {Pustovoit}}, \ and\ \bibinfo {author} {\bibfnamefont {S.~M.}\ \bibnamefont
  {Bezrukov}},\ }\href@noop {} {\bibfield  {journal} {\bibinfo  {journal} {Phys
  Rev E Stat Nonlin Soft Matter Phys}\ }\textbf {\bibinfo {volume} {80}},\
  \bibinfo {pages} {020904} (\bibinfo {year} {2009})}\BibitemShut {NoStop}%
\bibitem [{\citenamefont {Zilman}\ \emph {et~al.}(2009)\citenamefont {Zilman},
  \citenamefont {Pearson},\ and\ \citenamefont {Bel}}]{Zilman2009PRL}%
  \BibitemOpen
  \bibfield  {author} {\bibinfo {author} {\bibfnamefont {A.}~\bibnamefont
  {Zilman}}, \bibinfo {author} {\bibfnamefont {J.}~\bibnamefont {Pearson}}, \
  and\ \bibinfo {author} {\bibfnamefont {G.}~\bibnamefont {Bel}},\ }\href@noop
  {} {\bibfield  {journal} {\bibinfo  {journal} {Phys Rev Lett}\ }\textbf
  {\bibinfo {volume} {103}},\ \bibinfo {pages} {128103} (\bibinfo {year}
  {2009})}\BibitemShut {NoStop}%
\bibitem [{\citenamefont {Zilman}\ \emph {et~al.}(2010)\citenamefont {Zilman},
  \citenamefont {Talia}, \citenamefont {Jovanovic-Talisman}, \citenamefont
  {Chait}, \citenamefont {Rout},\ and\ \citenamefont
  {Magnasco}}]{Zilman2010PloSComputBiol}%
  \BibitemOpen
  \bibfield  {author} {\bibinfo {author} {\bibfnamefont {A.}~\bibnamefont
  {Zilman}}, \bibinfo {author} {\bibfnamefont {S.~D.}\ \bibnamefont {Talia}},
  \bibinfo {author} {\bibfnamefont {T.}~\bibnamefont {Jovanovic-Talisman}},
  \bibinfo {author} {\bibfnamefont {B.~T.}\ \bibnamefont {Chait}}, \bibinfo
  {author} {\bibfnamefont {M.~P.}\ \bibnamefont {Rout}}, \ and\ \bibinfo
  {author} {\bibfnamefont {M.~O.}\ \bibnamefont {Magnasco}},\ }\href@noop {}
  {\bibfield  {journal} {\bibinfo  {journal} {PLoS Comput Biol}\ }\textbf
  {\bibinfo {volume} {6}},\ \bibinfo {pages} {e1000804} (\bibinfo {year}
  {2010})}\BibitemShut {NoStop}%
\bibitem [{\citenamefont {Zilman}\ and\ \citenamefont
  {Bel}(2010)}]{Zilman2010}%
  \BibitemOpen
  \bibfield  {author} {\bibinfo {author} {\bibfnamefont {A.}~\bibnamefont
  {Zilman}}\ and\ \bibinfo {author} {\bibfnamefont {G.}~\bibnamefont {Bel}},\
  }\href@noop {} {\bibfield  {journal} {\bibinfo  {journal} {{J Phys: Condens
  Matter}}\ }\textbf {\bibinfo {volume} {{22}}},\ \bibinfo {pages} {{454130}}
  (\bibinfo {year} {{2010}})}\BibitemShut {NoStop}%
\bibitem [{\citenamefont {Hardt}(1981)}]{Hardt}%
  \BibitemOpen
  \bibfield  {author} {\bibinfo {author} {\bibfnamefont {S.}~\bibnamefont
  {Hardt}},\ }\href@noop {} {\bibfield  {journal} {\bibinfo  {journal} {Bull
  Math Biol}\ ,\ \bibinfo {pages} {89 }} (\bibinfo {year} {1981})}\BibitemShut
  {NoStop}%
\bibitem [{\citenamefont {Nadler}\ and\ \citenamefont
  {Schulten}(1985)}]{Nadler85}%
  \BibitemOpen
  \bibfield  {author} {\bibinfo {author} {\bibfnamefont {W.}~\bibnamefont
  {Nadler}}\ and\ \bibinfo {author} {\bibfnamefont {K.}~\bibnamefont
  {Schulten}},\ }\href@noop {} {\bibfield  {journal} {\bibinfo  {journal} {J
  Chem Phys}\ }\textbf {\bibinfo {volume} {82}},\ \bibinfo {pages} {151}
  (\bibinfo {year} {1985})}\BibitemShut {NoStop}%
\end{thebibliography}%

\end{document}